\documentclass[aps,twocolumn,showpacs,superscriptaddress,preprintnumbers,floatfix,amsmath,amssymb,eufrak,table]{revtex4}
\usepackage{ulem}
\usepackage{color}
\usepackage{graphicx}  
\usepackage{appendix}
 \usepackage{epsfig}
\usepackage[colorlinks=true,linkcolor= red, citecolor = cyan, urlcolor = blue ]{hyperref}
\usepackage{epstopdf}
\usepackage{amsmath}
\usepackage{subfig}
\usepackage{comment}
\usepackage{float}
\usepackage{multirow}

\newcommand{\be}{\begin{equation}}
	\newcommand{\ee}{\end{equation}}
\newcommand{\ba}{\begin{eqnarray}}
	\newcommand{\ea}{\end{eqnarray}}

\begin{document}

\title{A systematic study of initial state quark energy loss in fixed target proton nucleus collisions}


\author{Sourav Kanti Giri}\email{sk.giri@vecc.gov.in}
\affiliation{Variable Energy Cyclotron Centre, 1/AF Bidhan Nagar, Kolkata 700 064, India}
\affiliation{Homi Bhabha National Institute, Mumbai - 400085, INDIA}
\author{Partha Pratim Bhaduri}\email{partha.bhaduri@vecc.gov.in}
\affiliation{Variable Energy Cyclotron Centre, 1/AF Bidhan Nagar, Kolkata 700 064, India}
\affiliation{Homi Bhabha National Institute, Mumbai - 400085, INDIA}\email{partha.bhaduri@vecc.gov.in}
\author{Biswarup Paul}\email{biswarup.p@vecc.gov.in}
\affiliation{Variable Energy Cyclotron Centre, 1/AF Bidhan Nagar, Kolkata 700 064, India}
\author{Santosh K. Das}\email{santosh@iitgoa.ac.in}
\affiliation{School of Physical Sciences, Indian Institute of Technology Goa, Ponda-403401, Goa, India}%
\date{\today}

\begin{abstract}
 In this article, we investigate parton energy loss in cold nuclear matter by studying the ratio of Drell-Yan production cross sections in fixed-target proton-nucleus (p + A) collisions. We analyze Drell-Yan production cross-section data from the Fermilab E866 and E906 experiments using two different quark energy loss parametrization models and various parton distribution functions for 800 GeV and 120 GeV proton beams incident on light and heavy nuclear targets. The sensitivity of the energy loss parameter on the employed parton distribution function has been thoroughly investigated. Our results have been used to predict the target mass dependence of Drell-Yan production in upcoming proton-induced collisions at SPS and FAIR.
  
\end{abstract}
\maketitle
\section{Introduction}

The study of parton energy loss in high energy collisions has garnered considerable theoretical as well as experimental attention over past three decades. 
Experiments at RHIC and LHC collected plethora of data on jet-quenching which clearly indicate the hard partons suffering energy loss during their passage through the hot and dense deconfined medium \linebreak ~\cite{Bjorken:1982tu,Lee:2013bka,Cao:2020wlm,Cunqueiro:2021wls}. However, a quantitative understanding of the partonic energy loss in the plasma requires precise estimation of the same in the color confined cold nuclear matter. 
 Although it is widely accepted that energy loss in a hot medium is greater than in a cold medium, there is no consensus on the exact value of partonic energy loss per unit length or nuclear stopping power, dE/dx, in either case. The uncertainty in  determining the in-medium specific energy loss of partons primarily arises from the limitations of direct experimental measurements and the absence of widely accepted mechanisms and processes for its estimation.

In literature, two different experimental processes, \linebreak namely the semi-inclusive deep inelastic scattering (DIS) of lepton on nuclei and the Drell-Yan (DY) reaction in hadron-nucleus collisions~\cite{Drell:1970wh} have been identified as suitable tools to probe the energy loss suffered by the fast quarks via multiple scattering and gluon radiation as they traverse through nuclear medium~\cite{Arleo:2002ki}. The DIS data are usually analyzed to estimate the energy loss of the outgoing quark in the nuclear medium~\cite{Arleo:2003jz,  Song:2010zza}. 
On the other hand the DY reaction in hadron-nucleus collisions - $q + \bar{q} \rightarrow \gamma^{*} \rightarrow l^{+} + l^{-}$ at leading order (LO) - is considered an ideal probe to estimate the energy loss of incident quarks in the cold nuclear medium. As the final state lepton pairs do not interact strongly with the nuclear medium, the nuclear DY process offers a relatively clean experimental signal to probe the initial state effects. A comparative study of initial state and final state energy loss in nuclear matter~\cite{Vitev:2007ve} has shown that for sufficiently energetic partons the two mechanisms exhibit strikingly different dependencies in path length and parent parton energy. However, a definitive distinction of shadowing effects and parton energy loss in nuclear DY reactions has been found to be challenging. 
The Fermilab experiments E772~\cite{Alde:1990im} and E866~\cite{NuSea:1999egr} studied the nuclear dependence of the muon pair production in proton-induced DY reactions with 800 GeV proton beam. 
 The first measurement of energy loss by analyzing cross section ratios of Fe/Be and W/Be was conducted in Ref.~\cite{NuSea:1999egr}.
 As a large fraction of the data was collected for target quark momentum fraction, $x_{2} < 0.05$, hence EKS98~\cite{Eskola:1998iy} nuclear parton density distribution (nPDF) scheme was adopted to correct the shadowing effect. After compensating the suppression from nuclear shadowing, three different phenomenological parametrizations namely GM scheme~\cite{Gavin:1991qk}, BH scheme~\cite{Brodsky:1992nq} and BDMPS scheme~\cite{Baier:1996sk} having different kinematic dependency were used to extract the incident quark energy loss. A very small specific energy loss ($<0.44$ GeV/fm) consistent also with no energy loss scenario was observed. 
The small value of energy loss was attributed to the employed EKS98 nPDF set which had used the E772 DY data to determine the sea quark shadowing without accounting for parton energy loss effects. Subsequently in Refs.~\cite{Johnson:2001xfa,FNALE772:2000fmo}, the DY data from E772 and E866 experiments were analyzed in target rest frame where the final state lepton pairs were originated from the decay of a heavy photon bremsstrahlunged by an incident fast charged quark. This formalism had considerably lesser degree of shadowing and thus led to greater parton energy loss ($>2$ GeV/fm). It was later pointed out in Ref.~\cite{Garvey:2002sn} that such large value in part could be attributed to smaller parton path length used in the analysis, as the incident beam protons were required to travel a certain distance (equivalent to mean free path $\sim 2.5$ fm) inside the target nucleus before liberating an energetic parton. However, this approach was questioned later because although it employed the shadowing correction that could match the DIS data reasonably well at small $x_{2}$, but it failed to reproduce the observed nuclear dependence in the region $x_{2} > 0.04$, where maximum of the E772 data and around $50 \%$ of E866 data were collected. In Ref.~\cite{Arleo:2002ph} DY data from NA3 experiment~\cite{NA3:1983ltt} in 150 GeV $\pi^{-}+A$ collision were analyzed together with E866 data. Shadowing corrections were implemented through the EKS98 nPDF scheme, whereas parton energy loss was incorporated through the BDMPS approach.
Small shadowing contribution to the DY process in $\pi^{-}+A$ collisions was found to result in a quark transport coefficient in cold nuclear matter $\Hat{q}= 0.047 \pm 0.035$ GeV$^2$/fm corresponding to a mean energy loss rate of $ 0.2 \pm 0.15$ GeV/fm for fast quarks in a large nucleus. Small value of the energy loss as compared to previous extraction was attributed to the low statistics of NA3 data and poorly parametrized parton distributions of pions~\cite{Garvey:2002sn}. Since then multiple investigations have been performed to extract the initial state quark energy loss by analyzing then available DY data from the p+A and $\pi^{-}$+A collisions~\cite{Duan:2005wj, Duan:2008qt, Song:2012zz,Song:2017wuh, Neufeld:2010dz}, using phenomenological models that combine various available parametrizations of nuclear parton densities with different schemes of parton energy loss. These studies confirmed the strong dependence of initial state quark energy loss on parton distribution function. Larger values of specific energy loss or correspondingly $\hat{q}$ (for BDMPS formalism) are obtained if the employed nPDF schemes were  constrained only from the DIS data. On the other hand, global analysis of nuclear parton distribution functions including the E772 DY data in absence of any energy loss effect overestimates the sea quark shadowing and hence making the impact of initial state parton energy loss negligible in nuclear DY reactions. As the initial state energy loss in QCD bears certain similarities to the induced electromagnetic bremsstrahlung, it was quantified through quark radiation length $X_{0}$ in the NVZ model~\cite{Neufeld:2010dz}. Analysis of the E772/E866 DY data at $\sqrt{s_{NN}}=38.8$ GeV in absence of any nuclear modification of parton densities, indicated that in cold nuclear matter, $X_{0}$ may lie between 30 fm  and 160 fm. One common observation emerged from all these studies is that the problem in reliable determination of the initial state parton energy loss and its kinematic dependence in cold nuclear matter from nuclear DY process largely originates from the obscurity introduced by the nuclear modification of the parton densities and limited accessible experimental data. To overcome this situation a new experiment addressing the energy loss measurements and covering a phase space where other cold nuclear matter (CNM) effects are expected to have minimal contributions was desired.

In Ref.~\cite{Garvey:2002sn,Neufeld:2010dz} the authors advocated for a focused set of DY measurement with low energy proton beams that offer a unique opportunity to isolate initial-state energy loss effects from nuclear shadowing. Lower beam energies enhance the effects of energy loss and access a region of $x_{2}$ where shadowing effects are negligible. Quantitative prediction of differential DY cross section ratios in p+W and p+D collisions, as a function of quark momentum fraction of the projectile proton, $x_{1}$ revealed a clear sensitivity to the underlying parton energy loss for incident proton energies of 50 GeV and 120 GeV. Target mass ($A$) dependence of DY data in different $x_{1}$ bins was further predicted to be able to distinguish between the linear and quadratic variation of mean quark energy loss with path length traversed inside the nuclear matter. Attenuation of DY cross section at large values of Feynman $x_{F}$ versus linear size  of the target nucleus in 120 GeV p+A collisions was also anticipated to help determine $X_{0}$ with much reduced uncertainty. Preliminary data on differential DY cross section ratios in 120 GeV p+A collisions are now available from the Fermilab E906/SeaQuest experiment~\cite{Ayuso:2020zht}. The data have been first analyzed in Ref.~\cite{Arleo:2018zjw} within BDMPS parton energy loss framework with analytic parametrization of quenching weights, embedded in a NLO DY cross section calculation, using EPPS16 nPDF scheme and $\hat{q}$ extracted from J/$\psi$ measurements in the fully coherent regime. The measured DY cross section turned out to be in clear disagreement with nuclear PDF effects alone and in good qualitative agreement with initial state parton energy loss. A direct comparison of the E906 and E866/E772 results indicated absence of $x_{2}$ scaling implying violation of QCD factorization in DY production in p+A collisions. Subsequently E906 DY measurements are analyzed in Ref.~\cite{Song:2020vhy, Song:2021mzt} in conjunction with the DY data from E866~\cite{NuSea:1999egr} , NA3~\cite{NA3:1983ltt} and NA10~\cite{NA10:1987hho} experiments at the NLO level using BDMPS energy loss scheme with SW and analytic parametrization of quenching weights. Calculations including incident quark energy loss are found to agree well with the low energy data, whereas the effect of incoming gluon energy loss embodied in the primary NLO Compton scattering sub-process is found to be minimal. Target mass, target parton momentum fraction $x_{2}$ and scale $Q^{2}$ dependence of $\hat{q}$  in proton induced DY collisions has been studied in Ref.~\cite{Xu:2022usl} which reveals a significant impact of the atomic mass on the constant factor $\hat{q_0}$ of the quark transport coefficient. All these analyses are based on BDMPS energy loss formalism with a quadratic path length dependence. This is in contrast to the result obtained in Ref.~\cite{Lin:2017} where a linear dependency of quark energy loss as predicted by NVZ model is indicated.
In short, the precise path length dependence of incident quark energy loss in nuclear matter has not been settled yet. The goal of the present study is to perform a systematic reanalysis of the available DY data from different fixed target p+A collisions to determine the incident quark energy loss in cold nuclear matter and its parametric dependence on the medium path length, $L_{A}$. Data from Fermilab experiments E866 and E906 are investigated for this purpose. Previous studies of DY production in $\pi^{-}+A$  collisions revealed smaller rate of parton energy loss as compared to $p+A$ collisions. The available data are associated with large statistical errors due to lower intensity of the secondary $\pi^{-}$ beam. On the phenomenological level, parton distributions in pion are less precisely known than in protons and also the $\pi-N$ cross section is only $60 \%$ that of $N-N$ and thus resulting a larger mean free path for the pions in nuclear medium and correspondingly shorter path length for the partons. Hence DY production from these two type of reactions should not be treated on equal footing. We refrain from using data from $\pi^{-}+A$ reactions in the present study. Instead of performing a global fit to all available data sets as attempted earlier, we make simultaneous fit to the available nuclear DY cross section ratios separately for each experiment to look for any possible beam energy dependence of the quark energy loss. The obtained results are utilized to predict the prospect of DY measurements in upcoming fixed target p+A collisions at J-PARC~\cite{Nagamiya:2006en}, SPS~\cite{Falco:2023hwb, Alocco:2024hvm} and FAIR~\cite{Bhaduri:2022cql}.

The remainder of the paper is organized as follows. In Section II, the basic theoretical formalism of the DY process is introduced, followed by a short summary of the data selected for the present analysis in Section III. The results we obtain are discussed in Section IV. Finally the summary and conclusions are presented in Section V.



\section{Nuclear Drell-Yan reaction}

 \begin{figure*}[htpb]
  \includegraphics[scale=0.15]{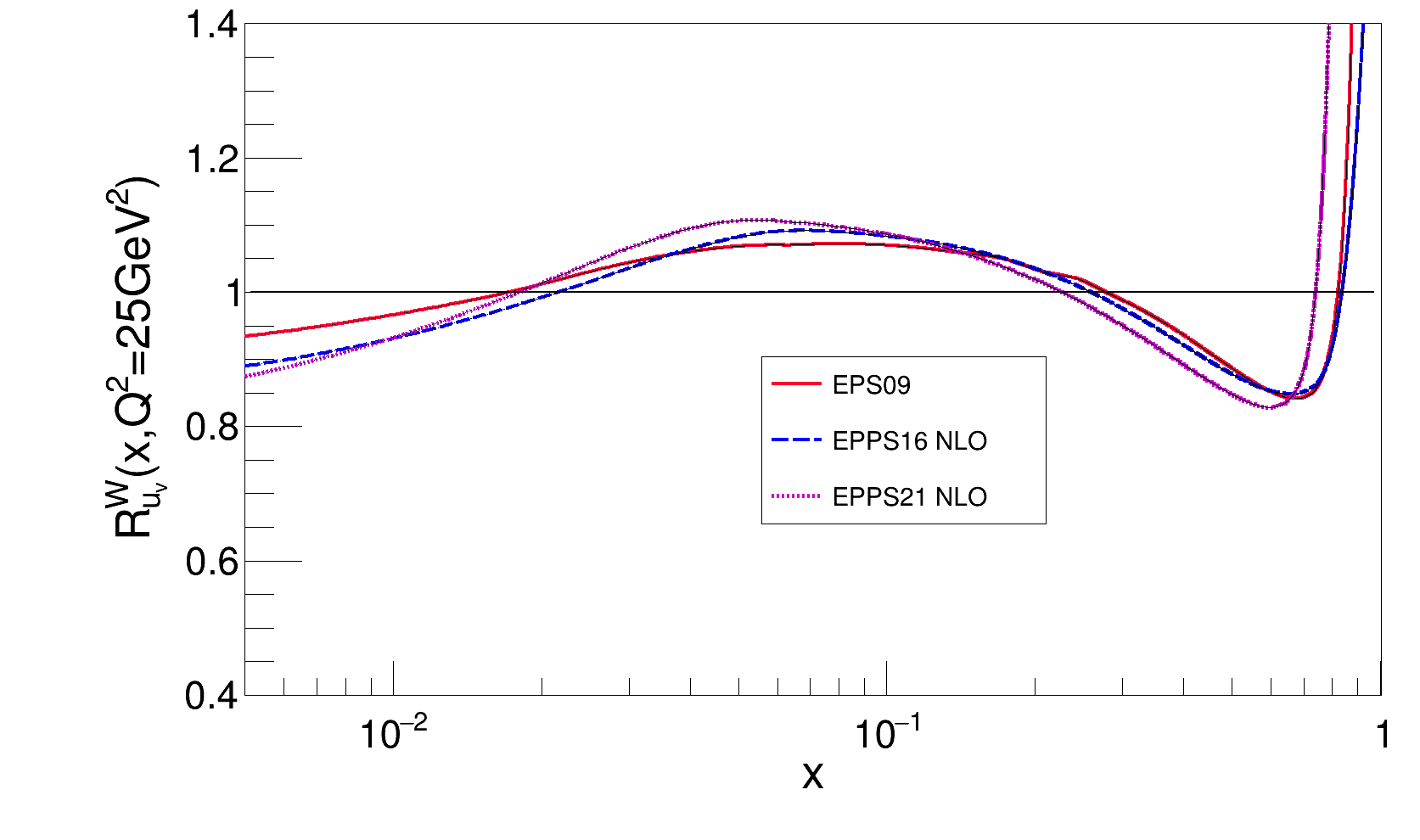}
    \includegraphics[scale=0.15]{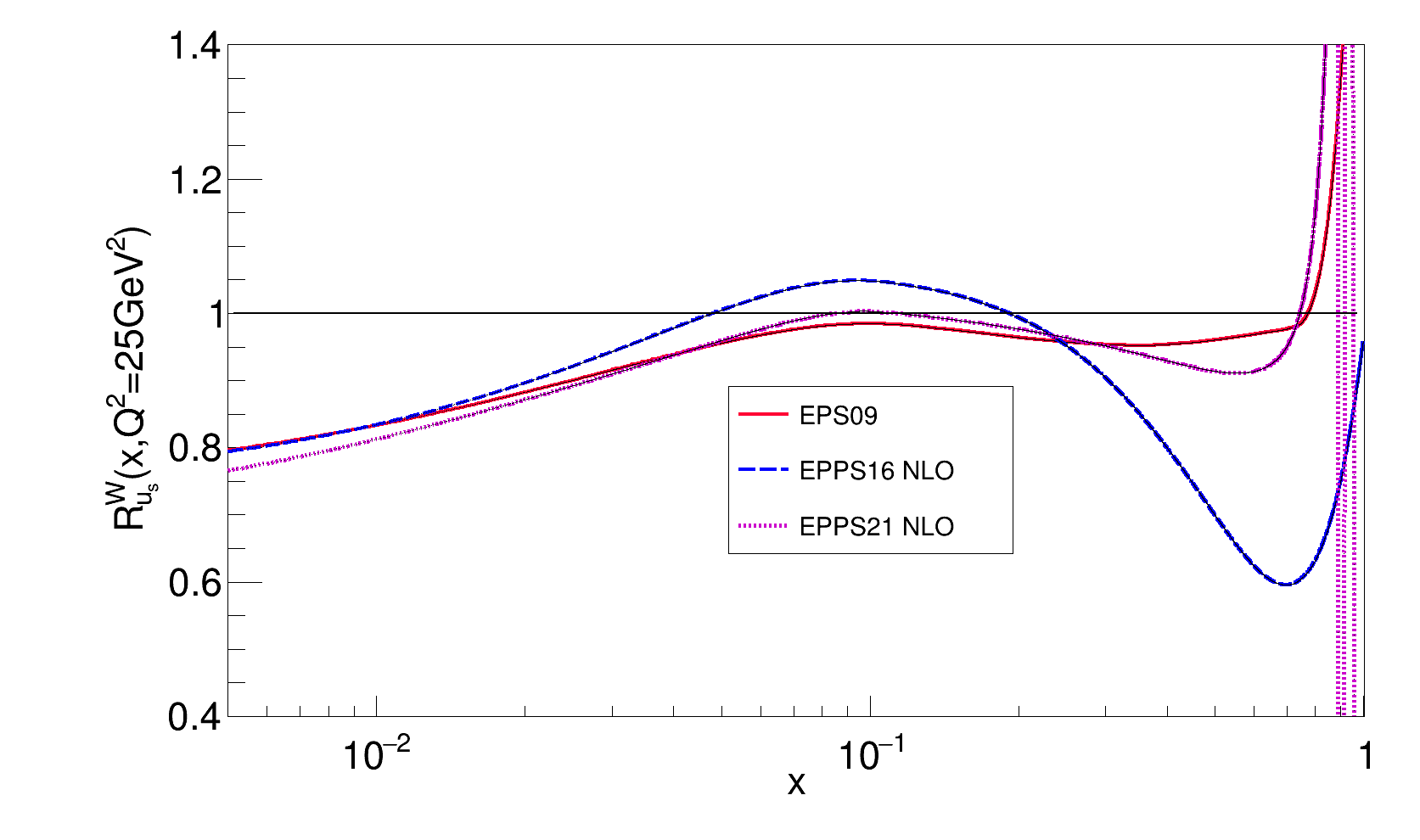}
    \includegraphics[scale=0.15]{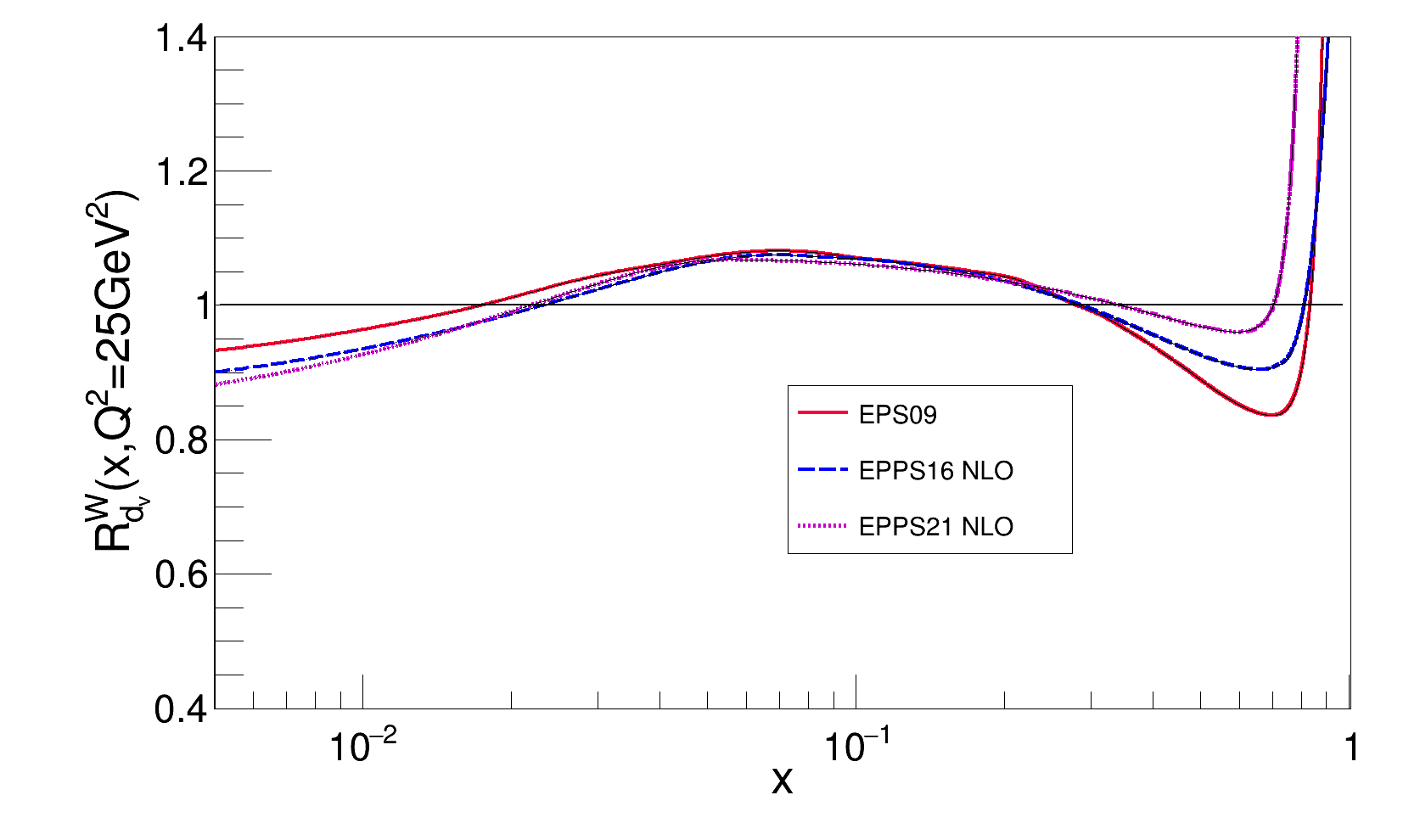}
    \includegraphics[scale=0.15]{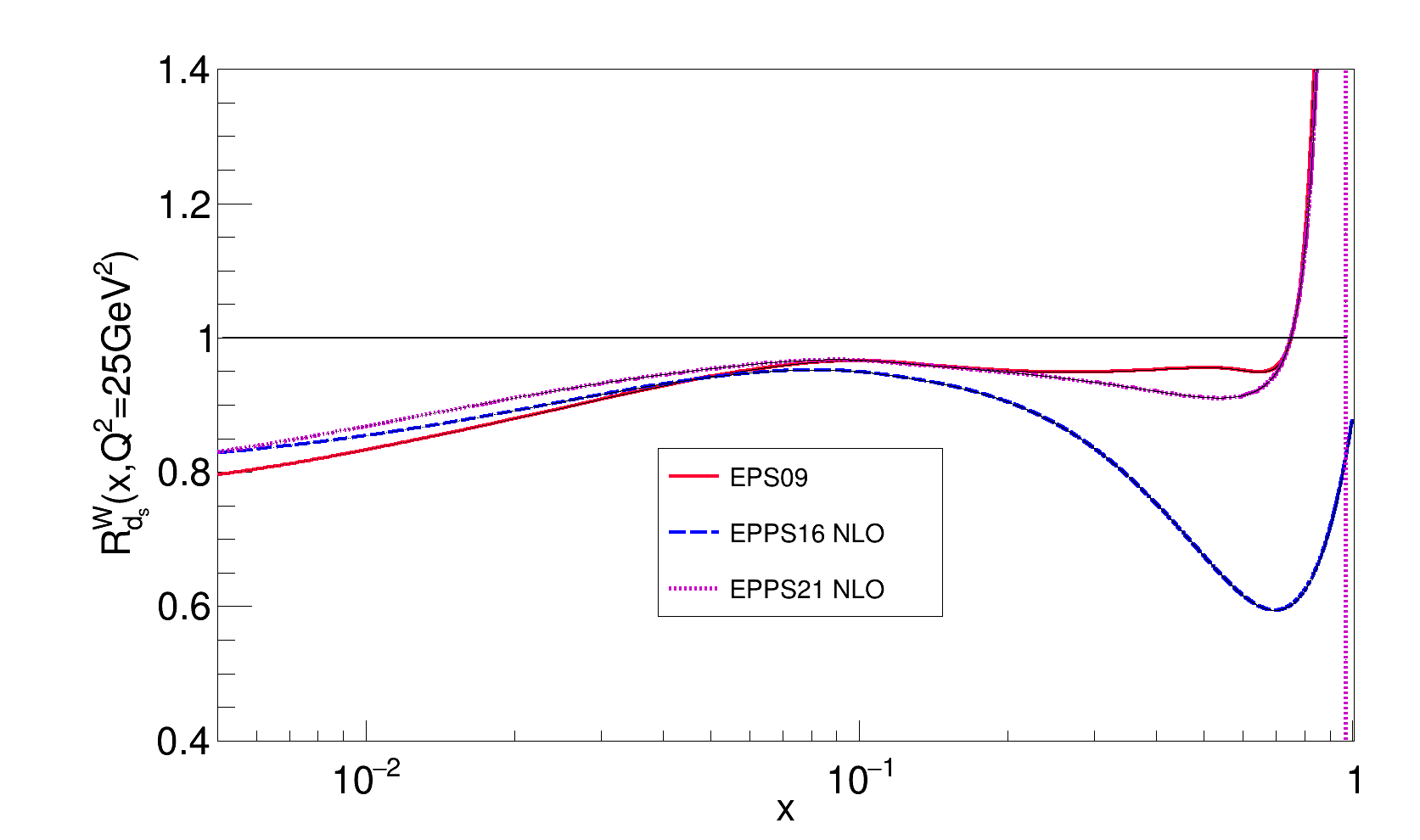}
    \caption{Variation of quark density distribution (expressed in terms of nuclear shadowing ratio, $R_{i}^{W}$) as a function of Bjorken variable $x$ evaluated at an interaction scale $Q^{2} = 25$ GeV$^{2}$, in tungsten (W) nucleus, for valence up quark $u_{v}$ (top left panel), sea up quark $u_{s}$ (top right panel), valence down quark $d_{v}$ (bottom left panel) and sea down quark $d_{s}$ (bottom right panel), as obtained from EPS09 LO, EPPS16 NLO and EPPS21 NLO nPDF schemes.}
     \label{fig:pdf}
 \end{figure*}

In the present work we are interested in the differential DY production cross section. 
The invariant mass M of the DY lepton pair is fixed by the center-of-mass energy of $q\bar{q}$ collision $\sqrt{\hat{s}}=(x_{1}x_{2}s)^{1/2}$ where $x_{1}$ and $x_{2}$ are the momentum fraction carried by the projectile and the target parton respectively and $\sqrt{s}$ denotes the center-of-mass energy of the hadronic collision. In absence of the incoming quark energy loss in cold nuclear matter, the LO double differential annihilation cross-section in the collinear factorization approach, in terms of Feynman scaling variable $x_{F}$, in the $p+A$ reactions is written as~\cite{Collins:1989gx, Gavin:1995ch} 
 \begin{eqnarray}
  \frac{d\sigma}{dx_{F}} &=K\frac{8\pi\alpha_{em}^{2}}{9s(x_{1}+x_{2})}\sum_{f} e_{f}^{2}\int \frac{dM}{M} \nonumber \\
  &\quad \times \Big{[}q_{f}^{p}(x_{1},M^{2}) {\bar q}_{f}^{A}(x_{2},M^{2}) \nonumber \\  
  &\quad +{\bar q}_{f}^{p}(x_{1},M^{2})q_{f}^{A}(x_{2},M^{2})\Big{]}
  \label{Eq:DY}
\end{eqnarray}
 where $K$ accounts for the higher order corrections, $\alpha_{em}$ denotes the fine structure constant. The sum is carried over\linebreak the light flavor ${\it{f} = u, d, s}$ of charge $e_{f}$. $q_{f}^{p(A)}(x,M^{2})$ and \linebreak $\bar{q}_{f}^{p(A)}(x, M^{2})$ denote the quark and anti-quark distribution function in the free proton (nucleon bound in the nucleus A) with Bjorken variable $x$ and virtual photon propagator mass square $M^{2}$. The integration over the pair mass $M$ depends on the kinematic coverage of the specific measurement. $x_{1}$and $x_{2}$ for a $2 \rightarrow 2$ process can be written as :
\begin{equation}
    x_{1}=\frac{1}{2}[\sqrt{x_{F}^{2}+4\frac{M^{2}}{s}}+x_{F}],   
\end{equation}
\begin{equation}
    x_{2}=\frac{1}{2}[\sqrt{x_{F}^{2}+4\frac{M^{2}}{s}}-x_{F}]
\end{equation}

such that $x_F = x_1 - x_2$. The quark density distribution inside a nucleus of atomic number $Z$ and mass number $A$ is expressed as the sum of the parton densities of the proton ($f_{i}^{p/A}$) and the neutron ($f_{i}^{n/A}$) as:
\begin{equation}
  \label{isospin}
  q_{f}^{A} = Z \ q_{f}^{p/A} + (A-Z) \ q_{f}^{n/A},
\end{equation}

where $f_{i}^{n/A}$ is obtained from $f_{i}^{p/A}$ by isospin conjugation: $u^{n/A}=d^{p/A}$, $d^{n/A}=u^{p/A}$, $s^{n/A}=s^{p/A}$. The DIS and DY measurements performed using nuclear targets have exhibited that the quark densities inside nuclei are significantly modified relative to those in free protons. Such nuclear modifications depend on $x$, $M^2$, and $A$.  Recent global fit extractions of nuclear parton distributions (nPDF) have been carried out by several different groups like DSZS~\cite{deFlorian:2011fp},\linebreak nCTEQ15~\cite{Kovarik:2015cma}, EPPS16~\cite{Eskola:2016oht} and EPPS21~\cite{Eskola:2021nhw} among the others. EPPS16 nPDF scheme first time used data from the LHC to better constraint the parton distributions in extended kinematic coverage compared to other available routines. In comparison to EPPS16, the recent EPPS21 nPDF scheme takes into account more data from p+Pb collisions at the LHC, namely 5 TeV double-differential CMS di-jet and \linebreak LHCb D-meson data, as well as 8 TeV CMS data on $ W^{\pm}$ boson production. Additionally the DIS measurements from Jefferson Lab probing nuclear PDFs at large and low virtualities, are also incorporated in EPPS21 analyses. In our present analysis we have thus used EPPS21 NLO nPDF \linebreak scheme to compute the DY production cross sections. 

\begin{figure*}[htpb]
     \includegraphics[scale=0.15]{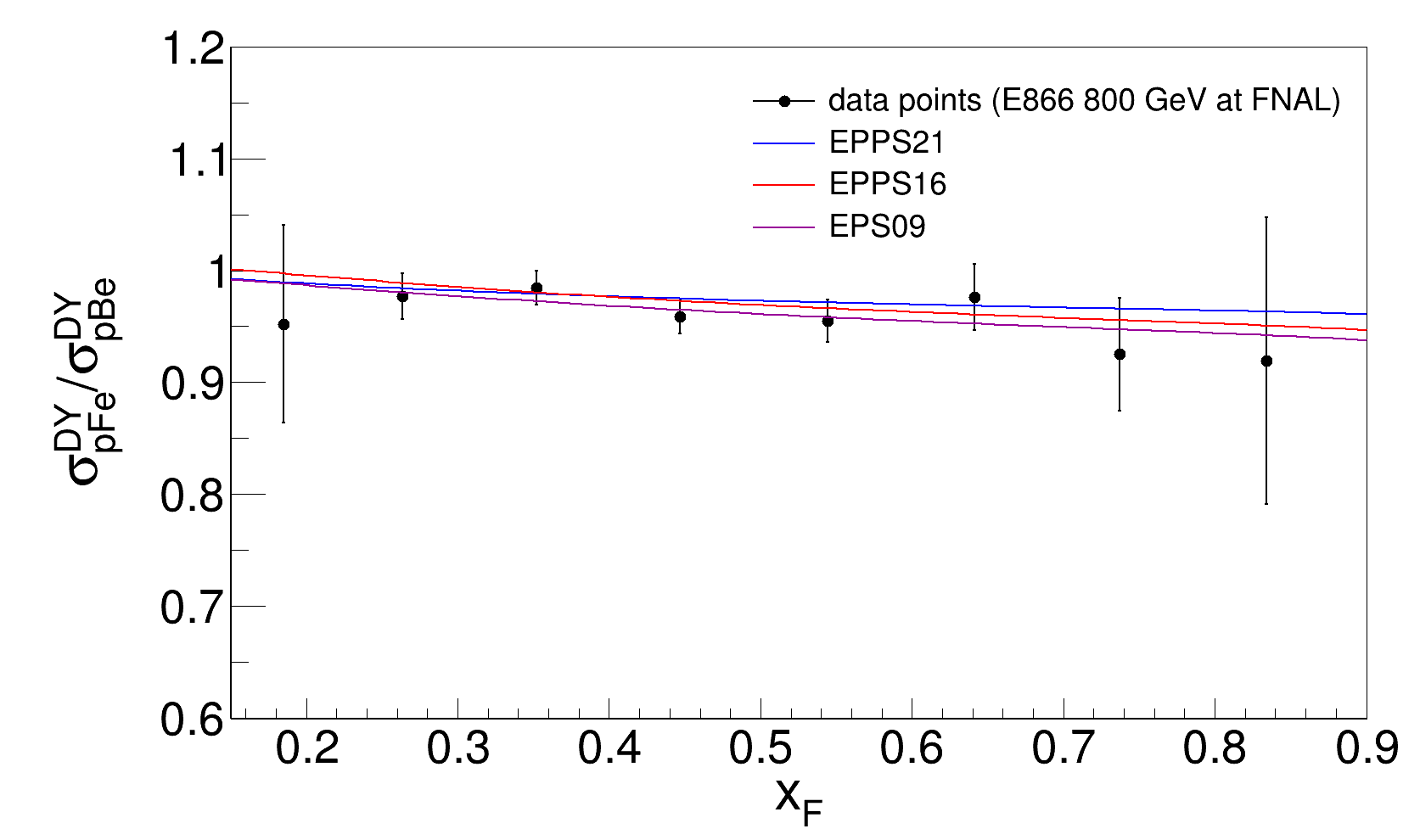}
     \includegraphics[scale=0.15]{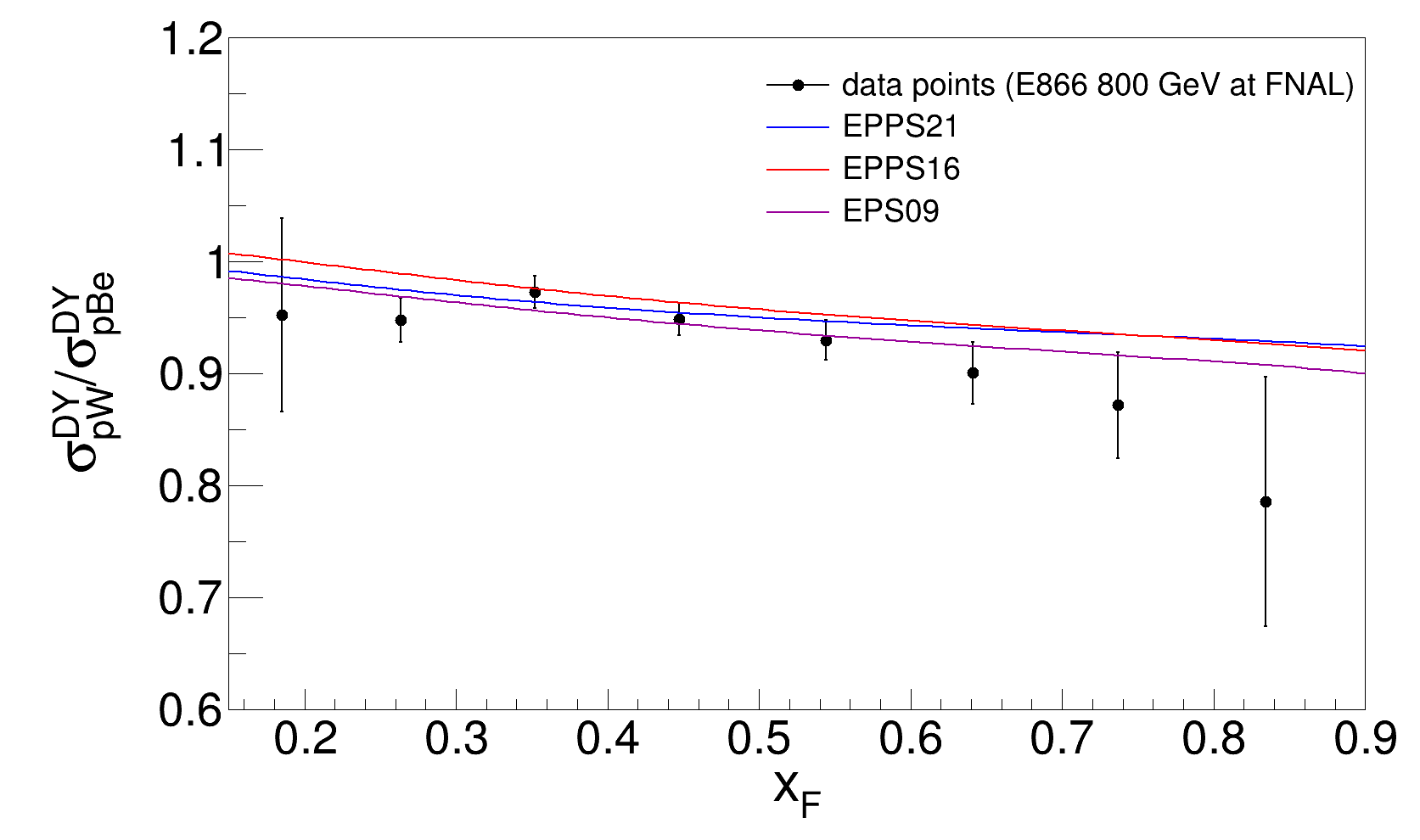}
     \includegraphics[scale=0.15]{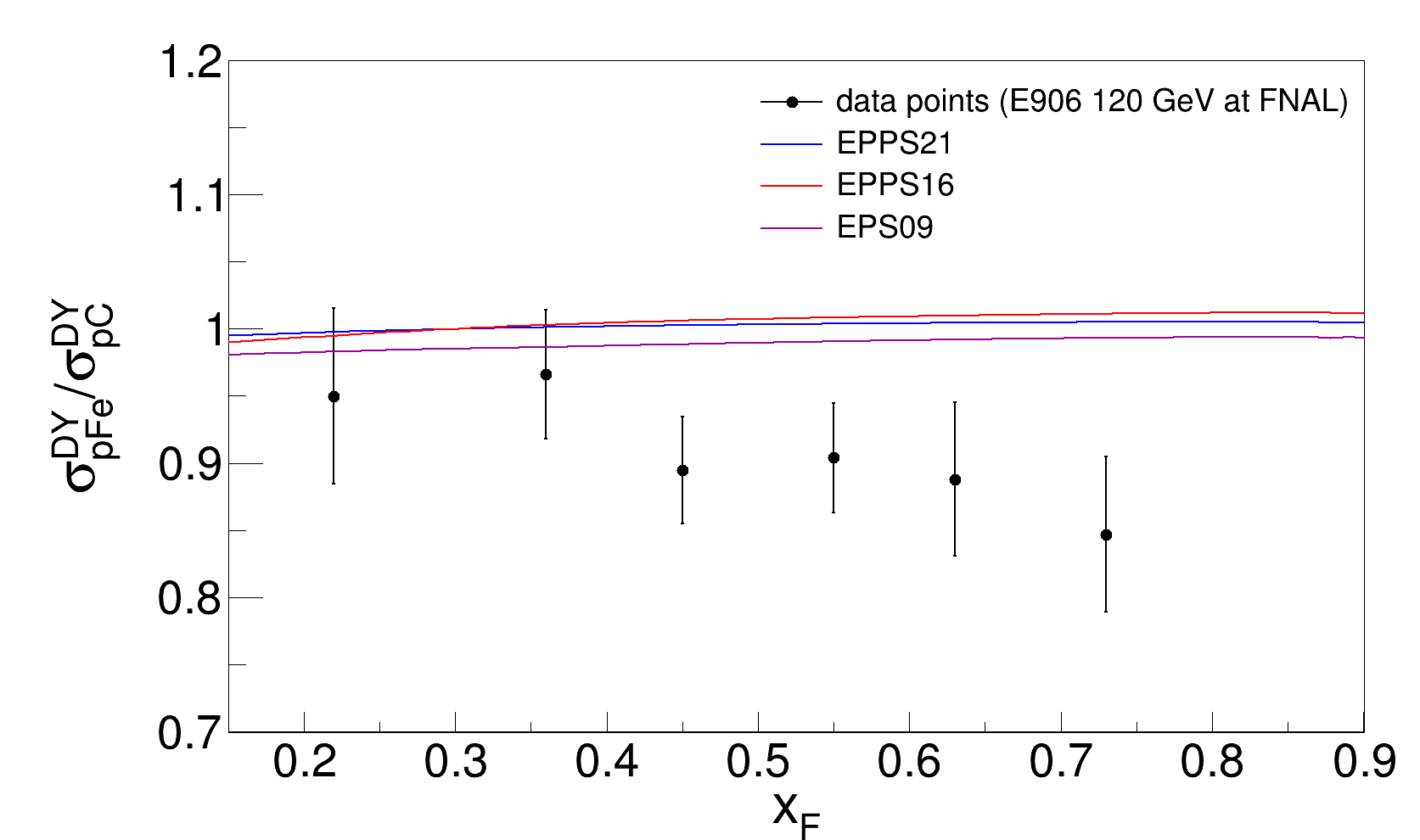}
     \includegraphics[scale=0.15]{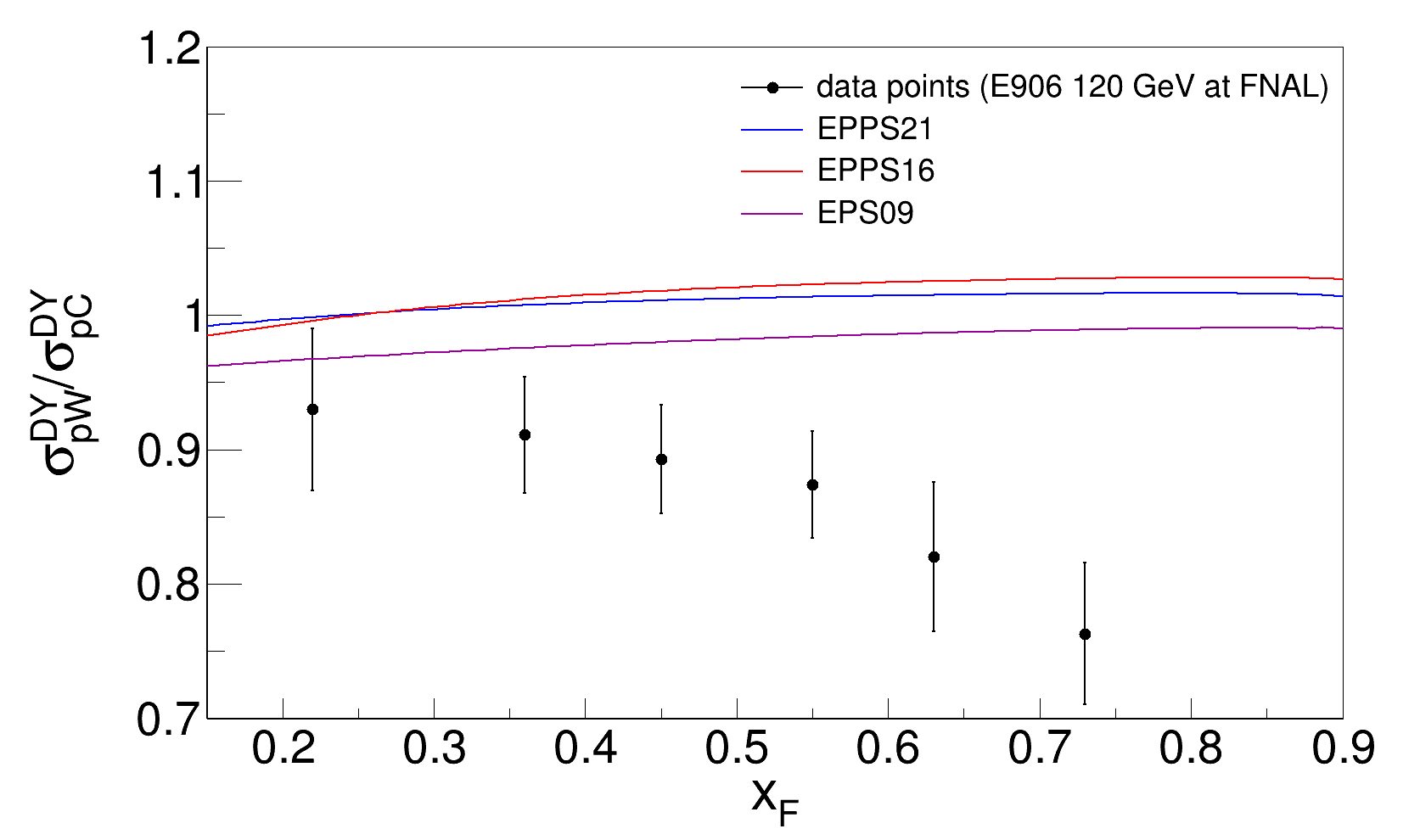}
    \caption{Nuclear Drell-Yan cross-section ratio as a function of $x_{F}$ as measured by the E866 (top panel) and E906 (bottom panel) experiments at Fermilab respectively in 800 GeV and 120 GeV p+A collisions. The solid curves are the calculated cross-section ratios computed for LO Drell-Yan process within the kinematic domain probed by these measurements, in absence of any beam quark energy loss but including the nuclear modification of quark densities inside the target nuclei modeled via EPS09 LO (magenta line), EPPS16 NLO (red line) and EPPS21 NLO (blue line) nPDF sets.}
    \label{fig:NoEloss}
 
\end{figure*}

\begin{figure}[htpb]
        \includegraphics[width=0.45\textwidth,height=5cm]{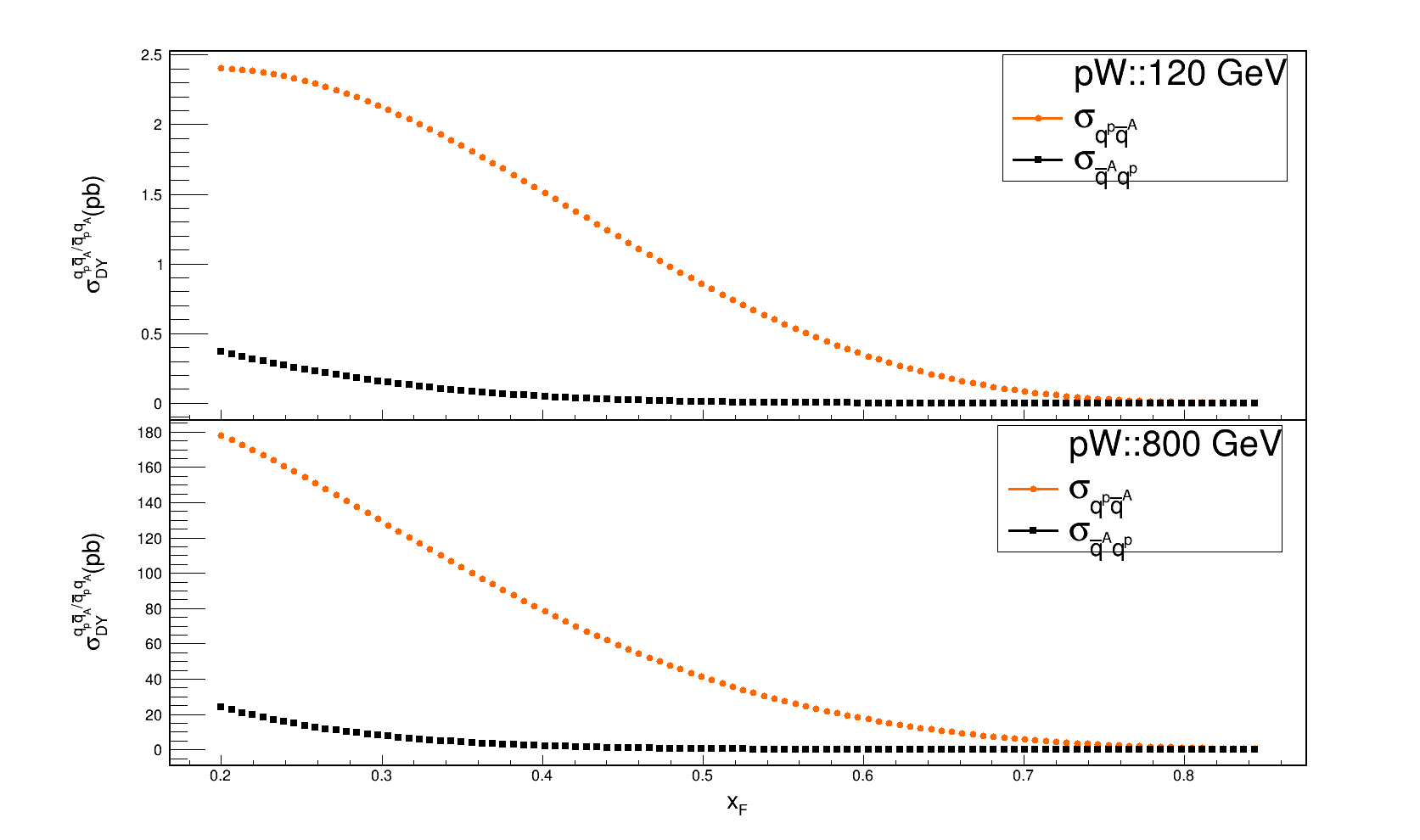}
        \caption{Variation of the leading order Drell-Yan production cross section from $q^{p}(x_{1})\bar{q}^{A}(x_{2})$ (red line) and $\bar{q}^{p}(x_{1})q^{A}(x_{2})$ (black line) annihilation processes as a function of $x_{F}$ in 120 GeV (top panel) and 800 GeV (bottom panel) p+W collisions.}
    \label{fig:DY_comp_xF}
\end{figure}

Now let us consider the energy loss scenario. In proton induced DY reactions on nuclei, in addition to nuclear modification of quark densities the incoming quark (antiquark) from the projectile may lose energy $\Delta E_{q}$ as it propagates through the nucleus. This energy loss occurs due to multiple scatterings with surrounding nucleons and soft gluon radiation before the hard $q\bar{q}$ annihilation process occurs. Incoming quark energy loss leads to an average change in its momentum fraction before the fusion, $\Delta x_{1} = \Delta E_{q}/E_{p}$, where $E_{p}$ is the incident proton beam energy in the target rest frame. As mentioned before, three parametrizations have been independently proposed in literature to model the fractional energy loss of the projectile quarks in cold nuclear matter of the target. In GM parametrization~\cite{Gavin:1991qk} the initial state quark energy loss is assumed to grow linearly with incident quark momentum fraction as: 
\begin{equation}
    \Delta x_{1} \approx \kappa x_{1} A^{1/3}
    \label{eloss_GM}
\end{equation}
As investigated in Ref.~\cite{Song:2012zz}, the available experimental data from $\pi^{-}+A$ collisions at lower beam energies rule out the possibility of GM energy loss. Within the BH~\cite{Brodsky:1992nq} formalism analogy to the photon Bremsstrahlung process in QED is used to obtain a form for the gluon radiation leading to a initial quark energy loss as:
\begin{equation}
    \Delta x_{1} \approx \frac{\alpha}{E_{h}} <L>_{A}
\end{equation}
where $\alpha$ signifies the incident quark specific energy loss in nuclear matter, $<L>_{A}={3 \over 4}R_{A}$ is the mean path length traveled by the incident quark inside the target nucleus of radius $R_{A} (\equiv R_{0}A^{1/3})$ having uniform matter density distribution. Finally the BDMPS~\cite{Baier:1996sk} approach is an extension of the BH model. Within this framework, the energy loss of sufficiently energetic partons is believed to depend on characteristic length and the transverse momentum ($p_{T}$) broadening of the parton. For finite size nucleus, both the factors vary as $A^{1/3}$ and hence the mean energy loss can be quantified as:

\begin{equation}
    \Delta x_{1} \approx  \frac{\beta}{E_{h}} <L>_{A}^{2}
    \label{eloss_BDMPS}
\end{equation}

Note that the BDMPS approach models the effects of initial state radiative energy loss in the Landau-Pomeranchuk-Migdal (LPM) regime corresponding to gluon formation time scales comparable to the medium length. The original formulation of BDMPS approach includes a distribution of induced energy loss $D(\epsilon)$ corresponding to the spectrum of radiated gluons. In the present analysis the quenching is approximated by shifting the projectile quark energy by a constant average energy loss $\Delta x_{1}$, on equal footing with other two models. In the BH and BDMPS formalisms, the average energy loss of the incident quarks depends linearly and quadratically on the traversed path length, respectively. Using these energy loss parametrization, it is possible to obtain the value of $\alpha$ and $\beta$ by analyzing the measured nuclear DY data on differential production cross section ratio. Considering the quark energy loss in the target nuclei, the multiple scattering shifts the incident quark (anti-quark) momentum fraction from $x_{1}^{'}=x_{1}+\Delta{x_{1}}$ to $x_{1}$ at the point of fusion. Hence target quark densities $q_f^A(x,M^2)$ have to be evaluated at ($x_{1} + \Delta{x_{1}}$) while computing nuclear DY production cross section. Due to steep behavior of the valence quark distributions at large $x_{1}$ even a small shift $\Delta{x_{1}}$ may generate substantial suppression in DY production in a heavier nucleus as compared to a light one. Taking both the nuclear shadowing and initial state quark energy loss into account, the single differential DY production cross section in $p+A$ collisions can be written as
\begin{eqnarray}
  \frac{d\sigma^{(pA)}}{dx_{F}} &=K\frac{8\pi\alpha^{2}}{9s(x_{1}+x_{2})}\sum_{f} e_{f}^{2}\int \frac{dM}{M} \nonumber \\
  &\quad \times \Big{[}q_{f}^{p}(x_{1}^{'},M^{2}) {\bar q}_{f}^{A}(x_{2},M^{2})  
\nonumber \\
  &\quad  + {\bar q}_{f}^{p}(x_{1}^{'},M^{2})q_{f}^{A}(x_{2},M^{2}\Big{]}
\end{eqnarray}

Both the shadowing effects and incoming quark energy loss lead to an $A$-dependent suppression of the DY production cross section. Shadowing attenuates the yield at small values of $x_{2}$ whereas the effects of energy loss are anticipated to be most pronounced at large values of $x_{1}$. As the acceptance of the fixed target detectors is generally biased to favor small values of $x_{2}$ in conjunction with large values of $x_{1}$, both effects are usually coupled in the measured data. In addition to these two effects, other process-dependent nuclear effects have been studied in the literature. 
 Since the present manuscript focuses on pair pT-integrated cross sections, the Cronin effect is not included. Dynamical shadowing arising from the coherent final state interactions of the recoil parton in the nuclear target also vanishes for LO DY process as no parton is present in the final state.~\footnote{Though there is a parton in the final state at NLO level, for typical DY dilepton masses ($M = 4 - 10$ GeV), the power-suppressed high-twist shadowing effects are expected to be negligibly small.}


\begin{figure*}[htpb]
    
    \includegraphics[scale=0.15]{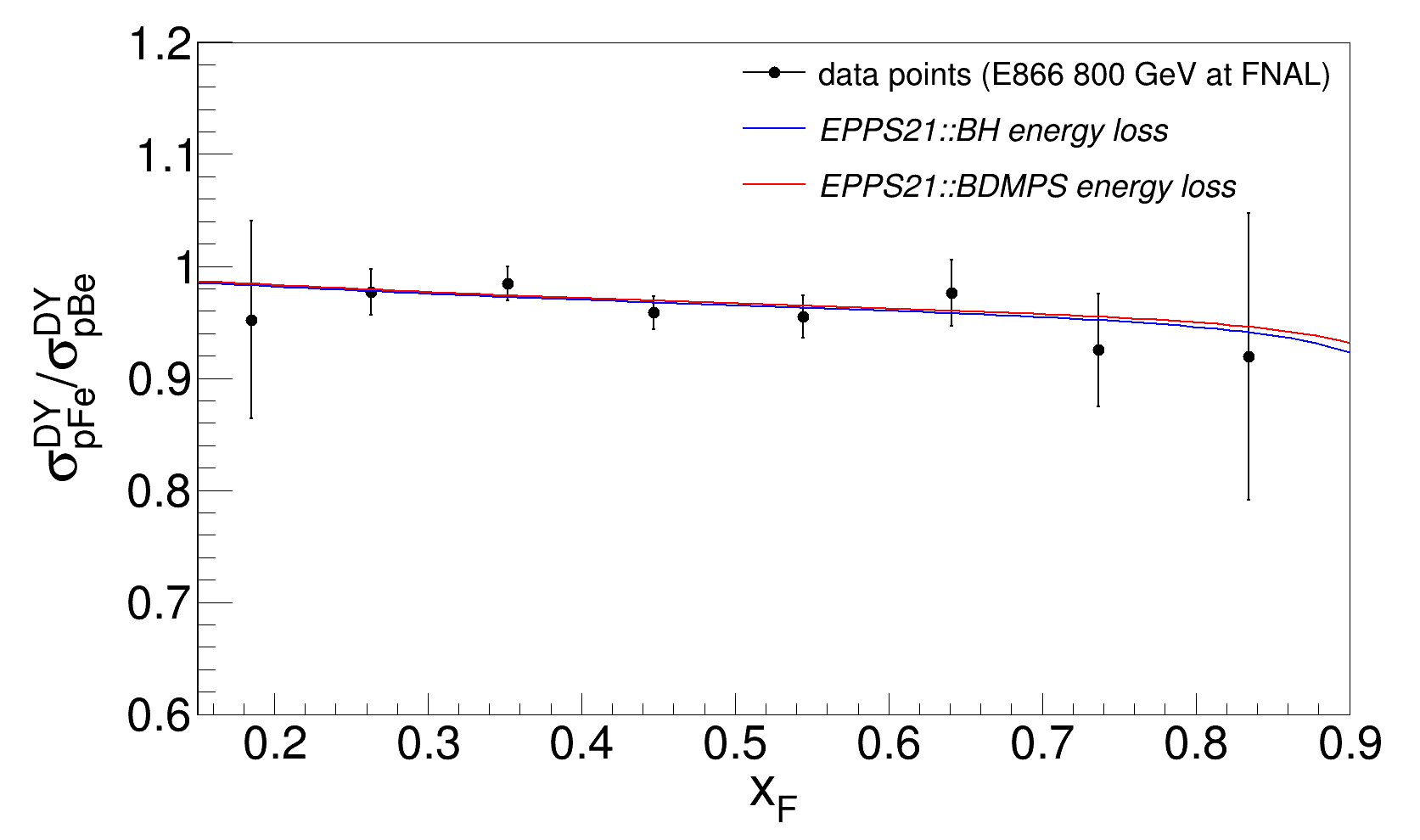}
    \includegraphics[scale=0.15]{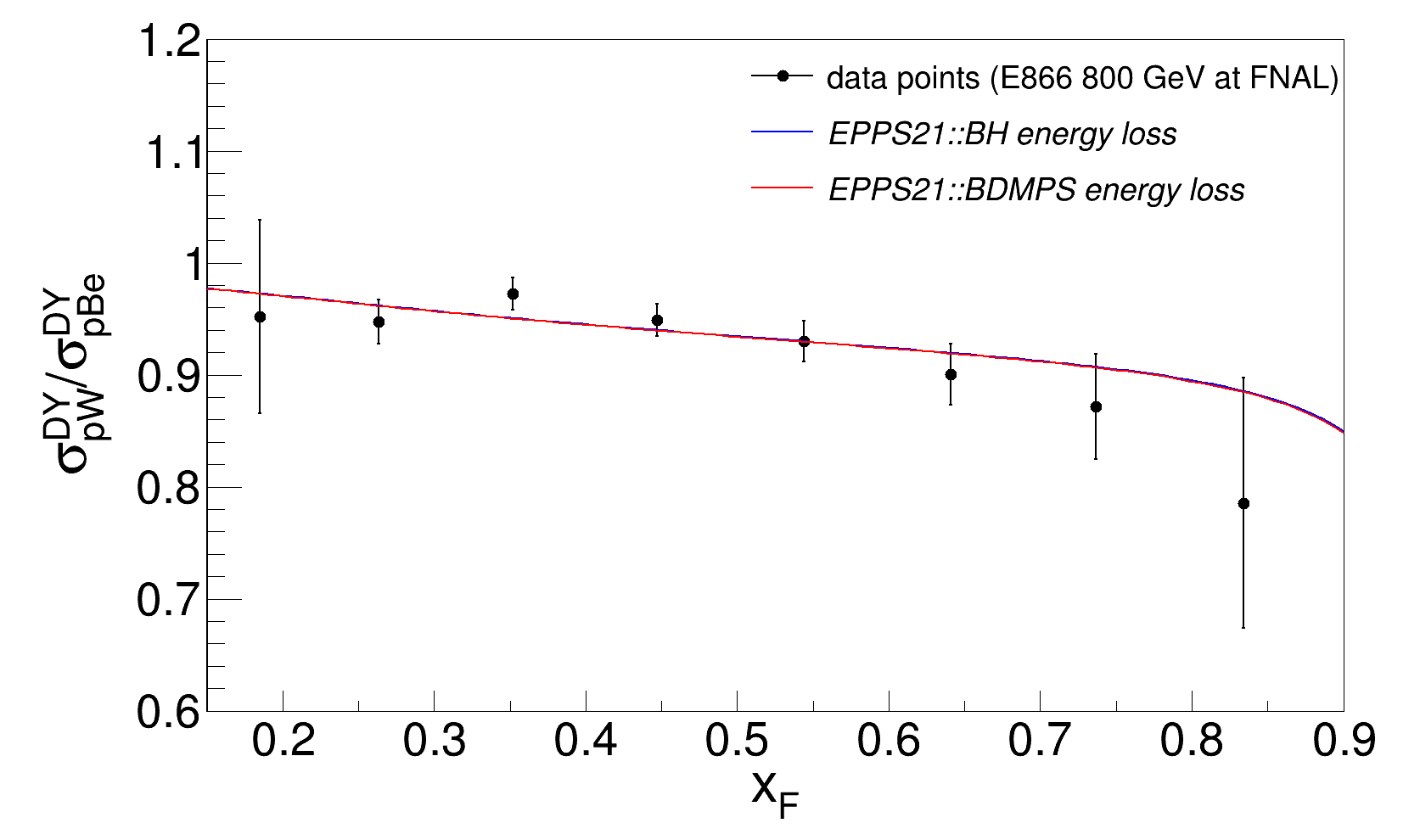}
    \includegraphics[scale=0.15]{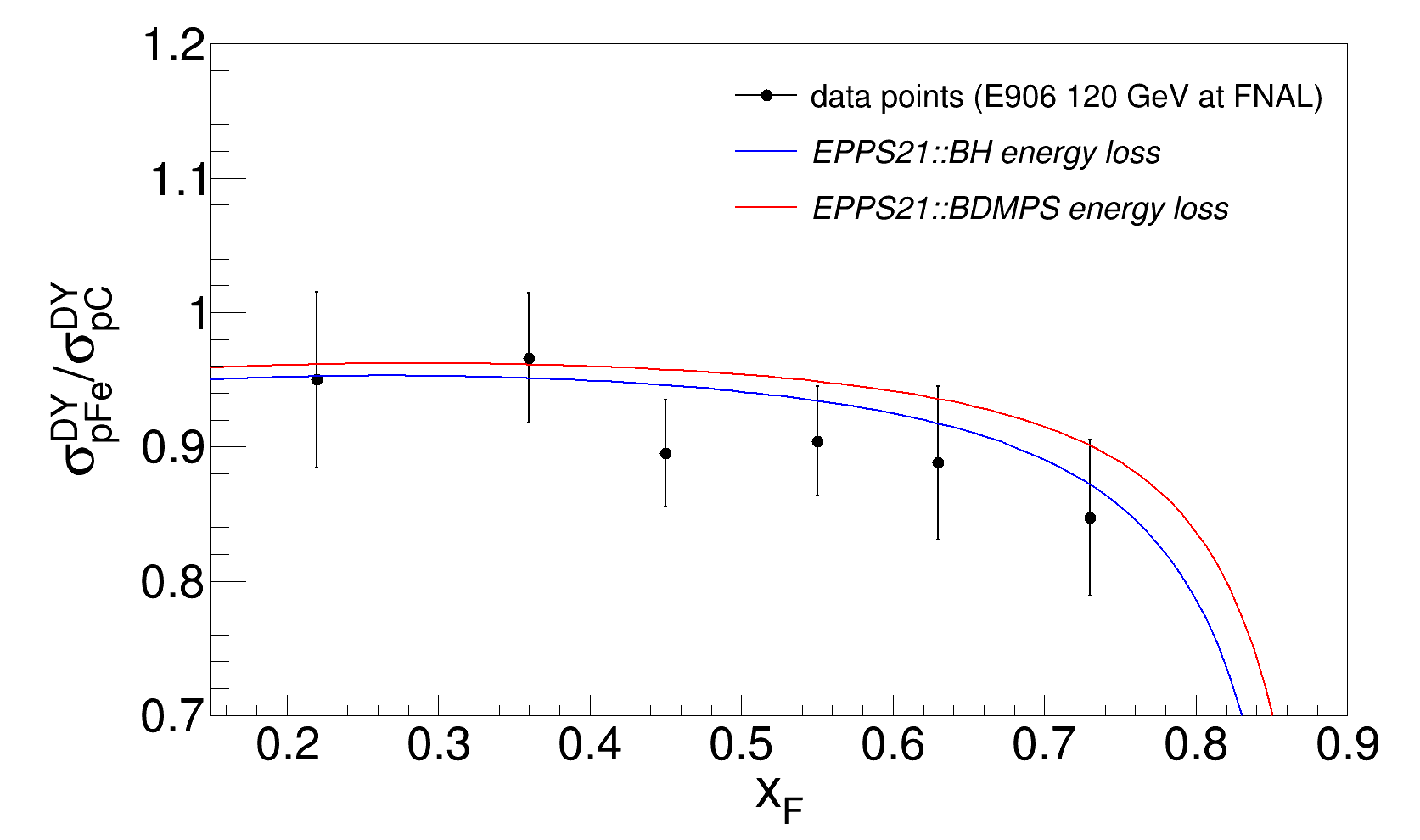}
    \includegraphics[scale=0.15]{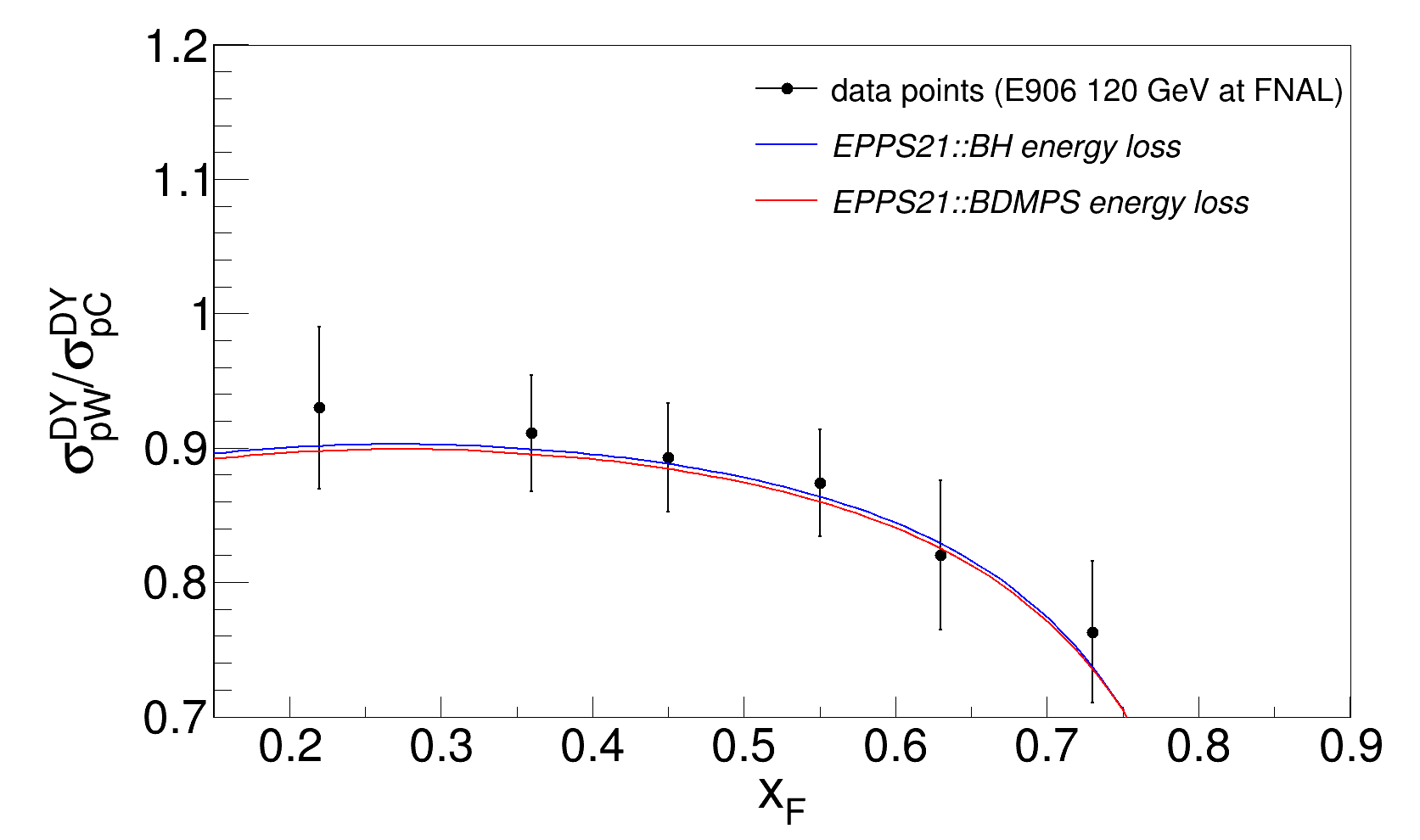}
    \caption{Nuclear Drell-Yan cross section ratios Fe/Be (Fe/C) (left) and W/Be (W/C) (right) versus $x_{F}$ as measured by the E866 (E906) experiments in 800 (120) GeV p+A collisions at Fermilab. The solid lines represent model calculations with best fit parameter, obtained from simultaneous fitting of DY cross section ratios for two heavy targets (Fe $\&$ W) at a given energy of the incident proton beam. Blue(Red) line represents BH(BDMPS) scheme of incident quark energy average energy loss. Nuclear modification of quark densities inside the targets are modeled using EPPS21 NLO PDF scheme.}
    \label{fig:E866}
\end{figure*}


\section{Experimental Data}

\begin{table*}

\centering

\begin{tabular}{|c|c|c|c|c|c|c|c|}
\cline{1-8}
experiment & $E_{Lab}(GeV)$ & Parametrization & PDF & $\alpha(GeV/fm) or \beta(GeV/fm^{2})$&$dE/dx(GeV/fm)$ & Probability & $\chi^{2}/ndf$ \\ 
\cline{1-8}
\multirow{6}{*}{E866} & \multirow{6}{*}{800} & \multirow{3}{*}{BH} & CT18ANLO&  1.715$\pm$0.199&1.72$\pm$0.20&0.97 & 0.40 \\ \cline{4-8}
 
                      &                         &                          & EPPS21NLO & 0.465$\pm$ 0.203&0.46$\pm$0.20& 0.95 & 0.48 \\ \cline{3-4} \cline{4-8}
                      &                         & \multirow{3}{*}{BDMPS} & CT18ANLO & 0.254$\pm$ 0.030 & 1.30$\pm$0.15  &0.93 & 0.50 \\ \cline{4-8}
                      &                         &                          & EPPS21NLO &0.068$\pm$ 0.030 & 0.35$\pm$0.15 &0.95  & 0.48 \\ \cline{1-3} \cline{4-8}
\multirow{6}{*}{E906} & \multirow{6}{*}{120} & \multirow{3}{*}{BH} & CT18ANLO & 0.437$\pm$ 0.053 &0.44$\pm$ 0.05& 0.98 & 0.32 \\ \cline{4-8}
        
                      &                         &                          & EPPS21NLO  & 0.467$\pm$ 0.053& 0.47$\pm$0.05&0.98 & 0.31 \\ \cline{3-4} \cline{4-8} 
                      &                         & \multirow{3}{*}{BDMPS} & CT18ANLO & 0.063$\pm$ 0.008 & 0.32$\pm$0.04 &0.86  & 0.56 \\ \cline{4-8}
                  
                      &                         &                          & EPPS21NLO &0.068$\pm$ 0.008 & 0.35$\pm$0.04& 0.86  & 0.56 \\ \hline

\end{tabular}

\caption{Summary of the best fit parameters obtained from simulataneous fitting of the Fe/Be (Fe/C) and W/Be (W/C) Drell-Yan cross section ratios as a function of $x_{F}$ measured by E866 (E906) collaboration at Fermilab in 800 (120) GeV p+A collisions. Results are obtained for BH and BDMPS parametrizations of initial state quark energy loss, with CT18ANLO free proton PDF and EPPS21 NLO nPDF schemes. For each case best fit value of the mean energy loss parameter ($\alpha$ or $\beta$), $p$-value and $\chi^{2}/ndf$ of the fit and the corresponding specific energy loss in W nucleus are tabulated.}
\label{tab:fitresult}
\end{table*}

\begin{figure}[htpb]
 \includegraphics[scale=0.15]{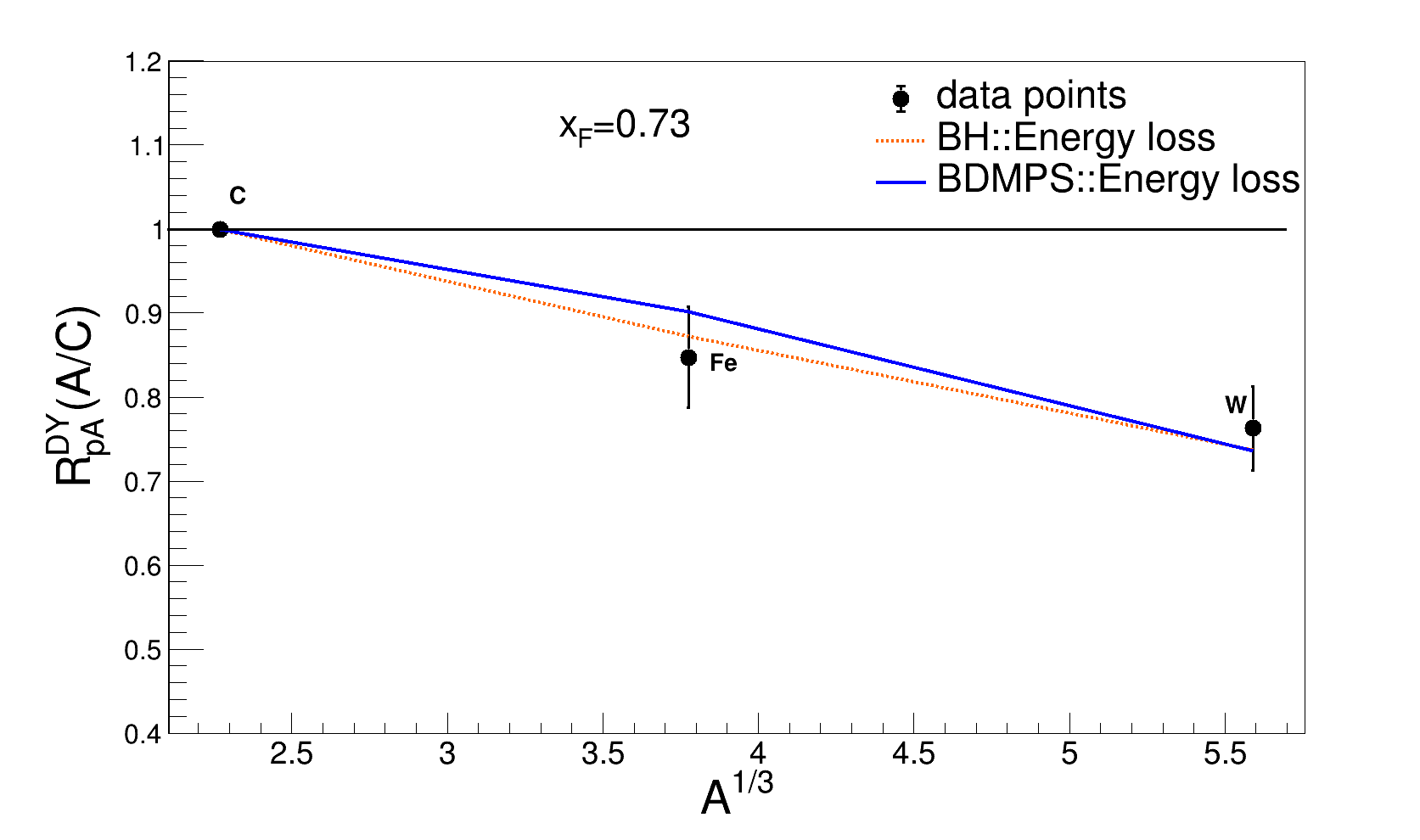}
\caption{The path length dependence of the energy loss effect. Variation of the ratio of the DY production cross sections in 120 GeV p+A collisions measured at the highest $x_{F} (~0.73$ as a function of $A^{1/3}$ for three different nuclear targets.} 
\label{fig:E906_LA}
\end{figure}

For the present work we have analyzed the the nuclear DY cross section ratios in p+A collisions as made available by E866~\cite{NuSea:1999egr} and E906~\cite{Ayuso:2020zht} collaborations at FermiLab. \linebreak E866/NuSea collaboration published their differential DY data in terms of W/Be and Fe/Be production cross section ratios as a function of $x_{1}$, $x_{2}$ and $x_{F}$ in 800 GeV p+A collisions. The DY muon pairs were collected within the mass range $4.0 < m_{\mu\mu} < 8.4$ GeV/$c^2$ covering the $x_{2}$ domain $0.01<x_{2}< 0.12$. Data from E906/SeaQuest experiment have been collected in 120 GeV p+A collisions, in the di-muon mass range $4.5 < m_{\mu\mu} < 5.5$ GeV/$c^{2}$. Results are available in terms of W/C and Fe/C differential DY production cross section ratios as a function of $x_{F}$, spanning over a $x_{2}$ regime $0.1 < x_{2} < 0.5$. To maintain consistency we have analyzed the $x_{F}$ dependent DY cross section ratios from both the experiments. The extracted values of energy loss parameter(s) are then used to compute and compare the integrated DY production cross section as a function of mass number for various target nuclei as measured in 400 GeV~\cite{NA50:2006rdp} and 450 GeV~\cite{NA50:2003pvd} p+A collisions by NA50 Collaboration at SPS, in the di-muon mass range $2.9 < m_{\mu\mu} < 4.5$ GeV/$c^{2}$. With 400 GeV proton beam, NA50 collected di-muons for six different nuclear targets (Be, Al, Cu, Ag, W, Pb) covering a phase space region $ -0.425 < y_{cms} < 0.33$. Data at 450 GeV were recorded for five different target nuclei (Be, Al, Cu, Ag, W) in the kinematic window $-0.5 < y_{cms} < 0.5$.


\section{Results and Discussions}

\begin{figure*}[htpb]
  \includegraphics[scale=0.15]{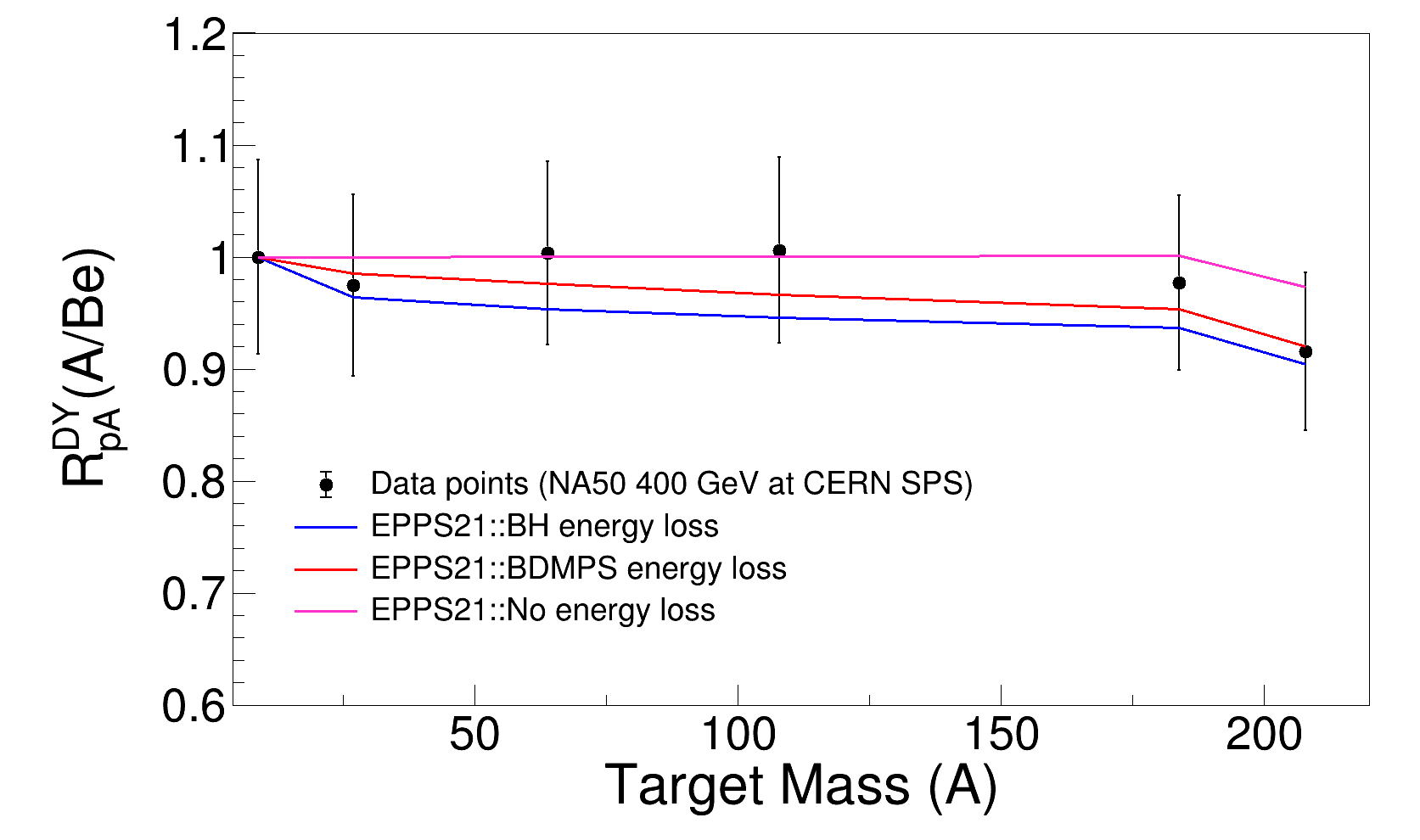}
  \includegraphics[scale=0.15]{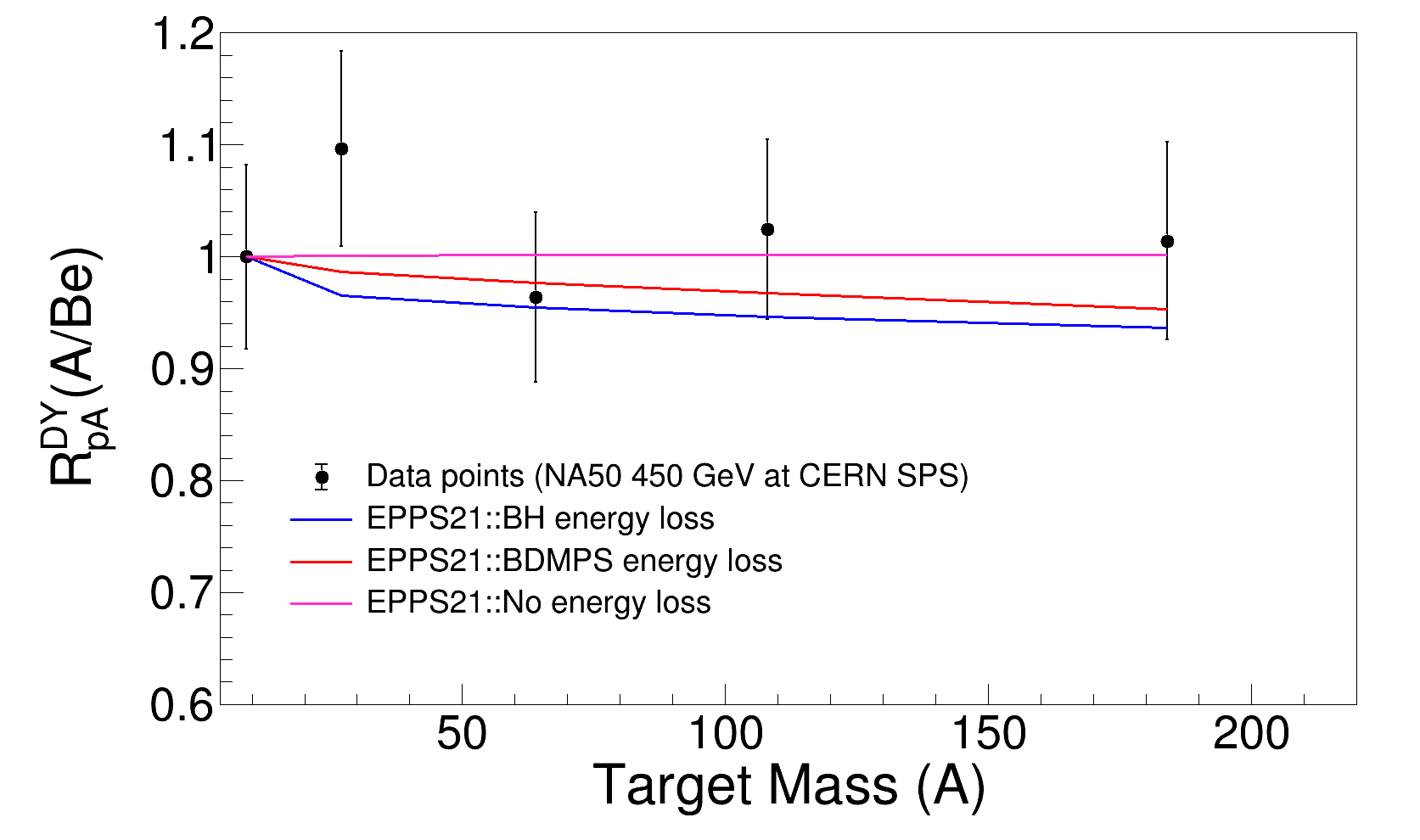}
  \caption{Integrated Drell-Yan (DY) cross section ratio in p+A collisions measured by NA50 experiment at CERN SPS for proton beam energy 400GeV (left) and 450 GeV (right) as a function of mass number (A) of the target. Solid curves are the predicted DY yield ratios for different targets for different scenarios corresponding to no incident quark energy loss (magenta) as well as quark energy loss following BH model (blue) and BDMPS model (red). EPPS21 nuclear PDF sets are used for all three cases to account for nuclear modification of target quark densities.}
  \label{fig:NA50}
\end{figure*}

In this section, we present and discuss the results of our analysis. We begin with the nuclear parton density distribution of valence and sea quarks in the heavy target nucleus. Fig.~\ref{fig:pdf} shows the so called shadowing ratio ($R_{i}^{W}$) signifying the variation of up and down quark densities in W nucleus, following EPPS21 NLO nPDF scheme, evaluated at an interaction scale $Q^{2} = 25$ GeV$^{2}$ relevant for the DY measurements under consideration, as a function of target $x_{2}$. For comparison, quark distributions from the previous variants of this routine namely EPS09 LO~\cite{Eskola:2009uj} and EPPS16 NLO packages are also included. For each case the central set with minimum uncertainty is chosen. As evident from the figure, valence quark ($u_{v}$  $\&$  $d_{v}$) distribution appear similar for 3 different parametrizations of nPDF with small to moderate anti-shadowing in the kinematic domain of E866 data ($0.01 <x_{2} < 0.12$), whereas anti-shadowing to eventual shadowing due to EMC effect in the kinematic range ($0.1 < x_{2} < 0.5$) probed by E906 experiment. On the other hand, for the opted nPDF parametrizations, sea quarks show stronger reduction in nuclear parton densities with EPPS16 scheme for large $x_{2}$, as compared to other two parametrizations.

\begin{figure*}
    \centering
    \includegraphics[width=0.49\linewidth,height=5cm]{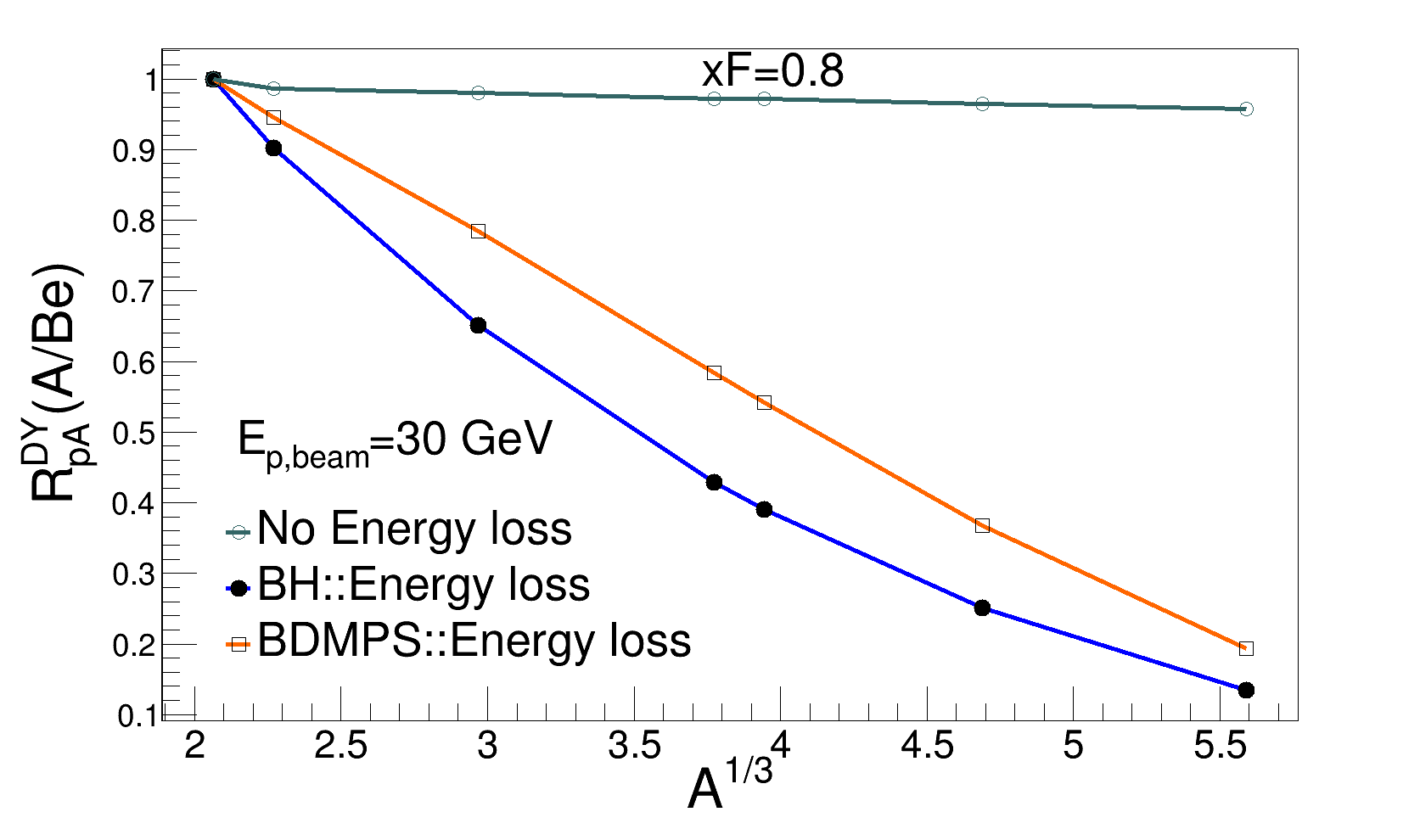}
     \includegraphics[width=0.49\linewidth,height=5cm]{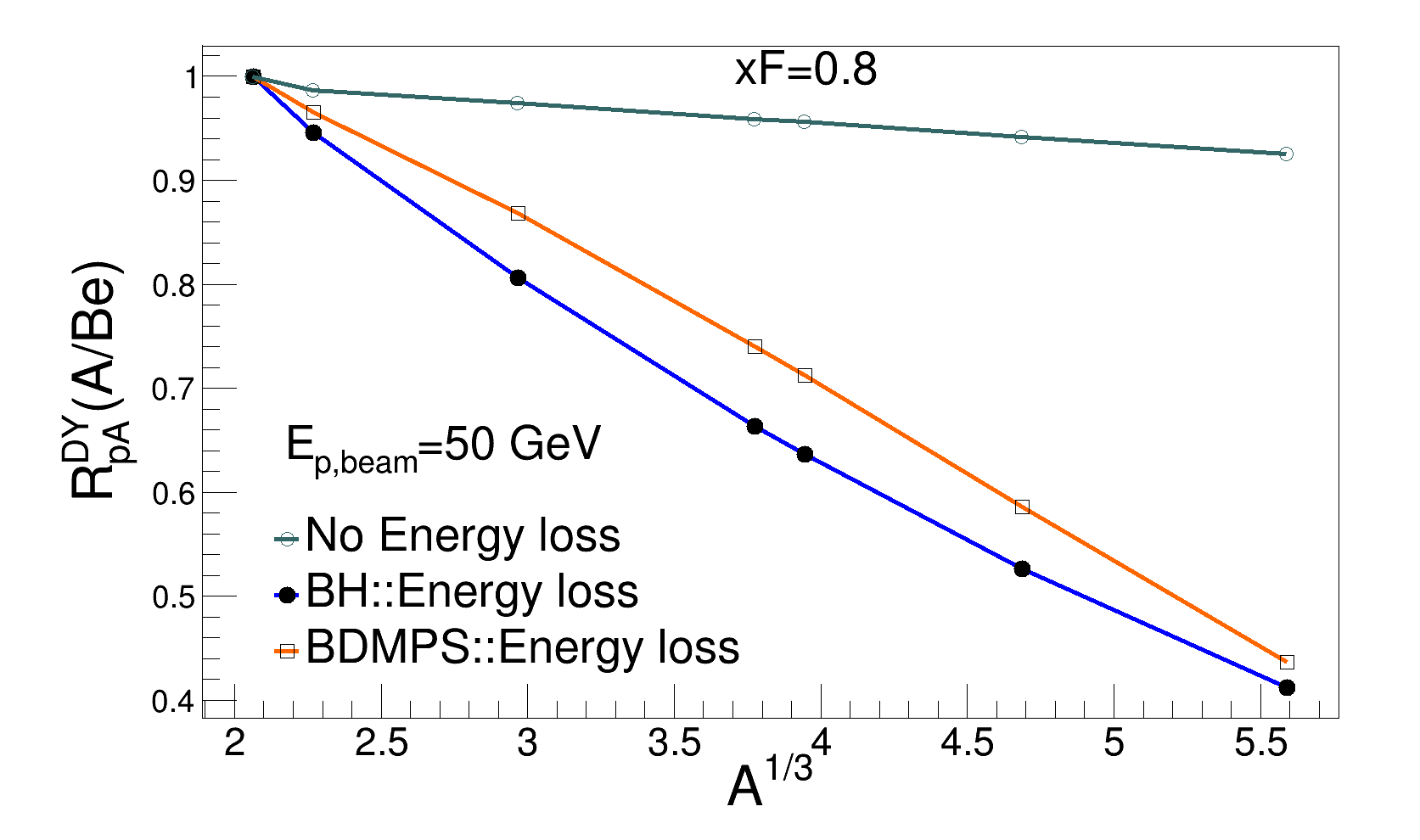}
    \includegraphics[width=0.49\linewidth,height=5cm]{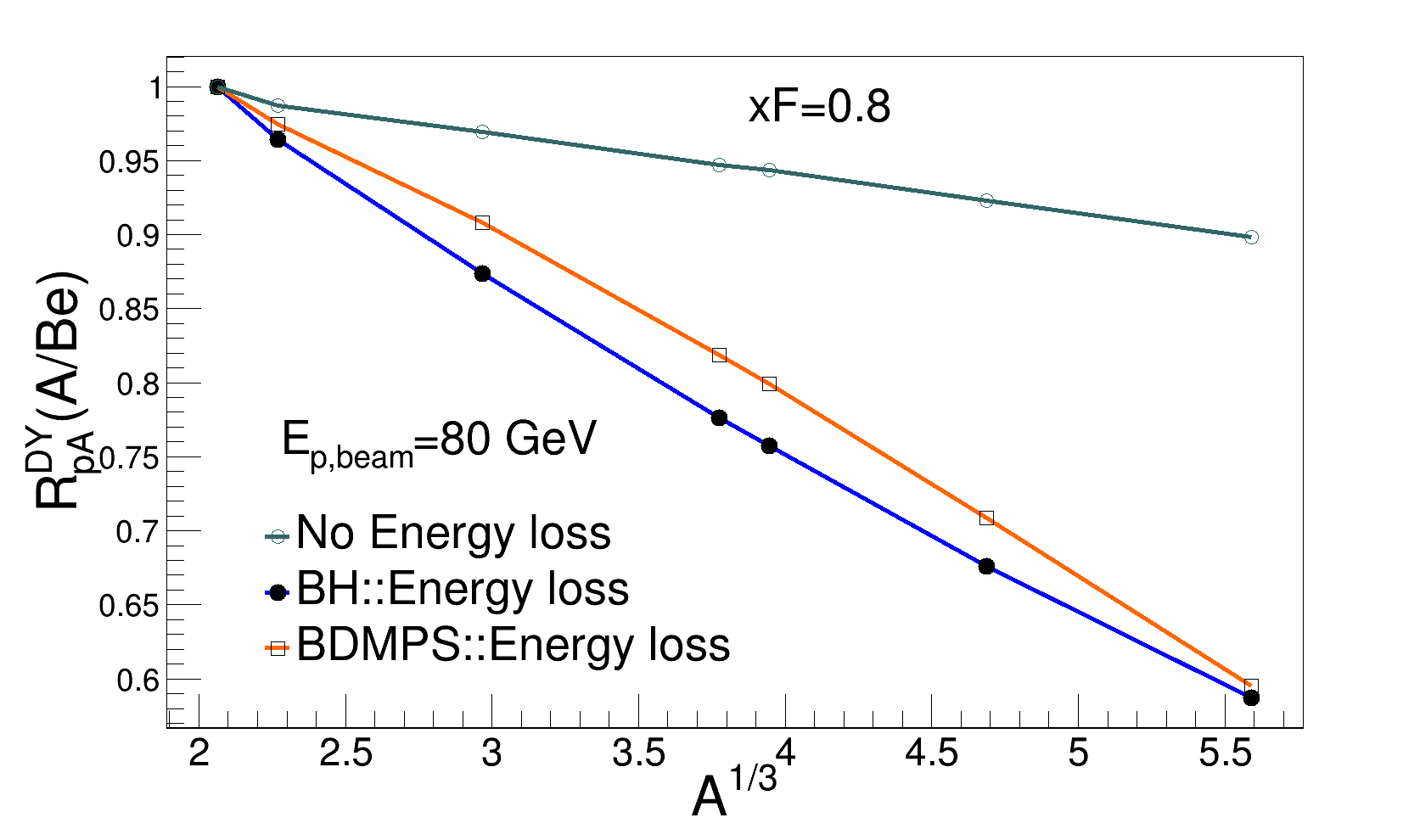}
        \caption{Theoretical predictions for the Drell-Yan di-muon cross section attenuation ($\sigma_{pA}/\sigma_{pBe}$) at large $x_{F}$ in 30 GeV, 50 GeV and 80 GeV p+A collisions, for different assumptions of the path-length dependence of initial state quark energy loss. Solid circles correspond to a linear dependence and the open squares correspond to a quadratic dependence of the mean partonic energy loss on the traversed path length inside the target nucleus.}
    \label{fig:Prospects}
\end{figure*}

Before delving into the analysis of energy loss, let us first look into the sole effect of nuclear parton densities on the nuclear DY production cross section. Fig.~\ref{fig:NoEloss} displays the Fe/Be (Fe/C) and W/Be (W/C) nuclear DY production cross section ratios as a function of $x_F$ at 800 (120) GeV, computed in the kinematic coverage of E866 (E906) experiment, in absence of energy loss, using the above three nPDF schemes. The experimental data points are also shown for comparison. For both the energies, the differential DY cross section ratios obtained from nuclear shadowing corrections alone look similar for all three nPDF parametrizations. This implies that difference in the shadowing ratio $R_{i}^{W}$ for sea quarks between EPPS16 NLO and other two nPDFs do not affect the computed DY production cross sections. The absolute value of quark densities are important while computing the cross sections and hence the effect of difference is minimal due to very small values of sea quark densities at large $x$. In line with the previous investigations performed, we see that at 800 GeV shadowing corrections alone can generate the required suppression in DY production at large $x_F$ as observed in the data, leaving a small room for energy loss effect to become operative. As noted in previous investigations, this is a common feature for all nPDF schemes including the presently adopted sets, where the sea quark distributions are parameterized using nuclear DY data from E772 and E866 experiments without accounting for any parton energy loss effects in cold nuclear matter. The scenario is different at 120 GeV. Shadowing corrections lead to almost a flat distribution of the DY cross section ratio in contrast to the large $x_F$ suppression observed in the data. It may be noted in this context that E906 data has not been used as input by the adopted nPDF sets for parametrising the parton distributions. The increasing trend in the computed cross section ratios at large $x_{F}$ as opposed to data results from the moderate anti-shadowing of the valence quarks in the low $x_{2} (x_2 \simeq 0.1)$ region probed by the measurements.

In this context, it is also interesting to note that the kinematic domain probed by the di-muon spectrometers of both E866 and E906 experiments correspond to forward acceptance collecting dileptons with high $x_{F}$ originating from beam parton of high $x_{1}$ and target parton of low $x_{2}$. The $x$ dependence of the parton densities indicates the dominant contribution to total DY production cross section would come from the $q^{p}(x_{1}, Q^{2})\bar{q}^{A}(x_{2}, Q^{2})$ term arising only from a beam quark annihilating with a target antiquark. For a quantitative estimation, we have shown in Fig.~\ref{fig:DY_comp_xF} the variation of the individual cross sections of the $q^{p}(x_{1}, Q^{2}){\bar{q}}^{A}(x_{2}, Q^{2})$ and \linebreak $\bar{q}^{p}(x_{1}, Q^{2})q^{A}(x_{2}, Q^{2})$ annihilation processes as a function of $x_{F}$ in 120 and 800 GeV p+W collisions. Since different nPDF sets generate similar effects on differential DY production cross section, henceforth we employ only latest EPPS21 NLO nPDF set for target partons and CT18ANLO~\cite{Xie:2021equ} package for free proton parton densities. As it appears at both the beam energies, the first term in Eq.\ref{Eq:DY} is much larger than the second term at small $x_{F}$ and the difference gradually decreases with increasing $x_{F}$ due to very small valence and sea quark densities of the projectile proton at large Bjorken-$x$. It may also be noted that the contribution due to up quarks ($u(x_{1})\bar{u(x_{2})}$) dominates over that due to down quarks \linebreak ($d(x_{1})\bar{d(x_{2})}$) because of $4 : 1$ charge-square weighting ratio and approximately $2 : 1$ $u$ to $d$ ratio in the incident proton at high $x_{1}$.

We now move on to analyze the differential DY cross section ratios for estimation of energy loss of the incoming quarks. 
Combining BH or BDMPS energy loss formalism with different PDF parametrizations the DY cross sections are calculated as a function of $x_{F}$ within the phase space window of the experimental measurements. For a quantitative evaluation of the effect of nuclear modification of parton densities on the extracted value of quark energy loss from data, in addition to the EPPS21 nPDF set we also use CT18ANLO free proton set for target partons to the calculate the DY cross section ratios. The energy loss parameter is extracted by simultaneously fitting the $x_{F}$ dependence of the Fe/Be(Fe/C) and W/Be(W/C) DY cross section ratios for E866(E906) data points. Fitting is performed with ROOT~\cite{root} software package using $\chi^{2}$ minimization~\cite{minuit2} technique. The best fit values of the parameters $\alpha$ and $\beta$ along with the $\chi^{2}$/ndf, $p$-value of the fit and the corresponding mean specific energy loss $<dE/dx>$ in W nucleus, for the two PDF sets are summarized in Table~\ref{tab:fitresult}. 
At 800 GeV, for either of the quark energy loss scenarios, the extracted energy loss values are much bigger for free proton PDF. This can be attributed to the presence of strong shadowing effects in Fe and W nuclei for the EPPS21 nPDF scheme, in the low $x_{2}$ region of E866 measurements. However, at 120 GeV the extracted energy loss parameters are similar for free proton pdf and nPDF, for either of the quark energy loss models. As seen above the shadowing corrections by EPPS21 nPDF scheme generate almost flat DY cross section ratio close to unity as a function of $x_{F}$. Hence, similar magnitudes of energy loss parameters are necessary to reproduce the  observed suppression present in the data. This also indicates the minimal dependence of the extracted energy loss on the underlying nPDF scheme in the phase space domain of E906 DY measurements. For the best fit values of $\alpha(\beta)$ in line with linear(quadratic) energy loss scenario, data-to-model comparison results are displayed in Fig.~\ref{fig:E866}. As evident from the figure, inclusion of incident quark energy loss is essential to match the observed suppression at high $x_{F}$ at 120 GeV beam energy. However, in contrast to the previous expectations, the present data set cannot distinguish between the linear and quadratic path length dependence of the mean energy loss in cold nuclear matter. 

Note that in case of BDMPS model, it is also possible to extract the mean value of quark transport co-efficient $\hat{q}$ in cold nuclear matter. Under the assumption that the energy of the radiated gluons inside the target nucleus is much smaller than the energy of the incident beam quark, the mean BDMPS incoming quark energy loss can be given by $ - \Delta E= {1 \over 2}\alpha_{s}C_{R}\hat{q}L_{A}^{2}$, where $C_{R} = 4/3$ is the Casimir (color) factor of the quark. The transport coefficient has been parametrized as~\cite{Arleo:2012rs}.
\begin{equation}
    \hat{q}(x) \equiv{  \hat{q}_{0} \left( \frac{10^{-2}}{x} \right)^{0.3}} ;\ x = \min(x_{0}, x_{2}) ;\ x_{0} \equiv{ \frac{1}{2m_{p}L}}
    \label{transportEqu}
\end{equation}
The typical value of $x_{0}$ for W nucleus comes around $x_{0} \simeq 0.02$. DY data measured by E866/E906 collaboration primarily probe region of $x_{2}$ where $x_{2} > x_{0}$. Thus following Eq.~\ref{transportEqu} transport coefficient is frozen at $x_{0}$, $\hat{q}({x_{0}}) \approx 0.8\hat{q}_{0}$. Freezing the strong coupling at $\alpha_s(M^{2}) = 0.5$ and comparing with Eq.~\ref{eloss_BDMPS} we get $\hat{q} = 6\beta$. This leads to mean value of $\hat{q} = 0.3 \pm 0.03 $ GeV$^2$/fm for CT18ANLO free proton pdf and $\hat{q} = 0.081 \pm 0.035$ GeV$^2$/fm for EPPS21 nPDF at 800 GeV. As previously mentioned, using EKS98 LO nPDF set $\hat{q}= 0.047 \pm 0.035$ GeV$^2$/fm was reported in~\cite{Arleo:2002ph}. Apart from the different variants of the nPDF scheme, the difference can mostly be ascribed to replacement of the distribution in the induced energy loss $D(\epsilon)$ by a constant mean value in the present case which was found to be larger than the actual loss contributing to quenching, particularly at large $x_{F}$, where DY production shows a sharp fall. From DY data of E906 experiment we extract the mean quark transport coefficient $\hat{q} \simeq 0.08 \pm 0.01$ GeV$^2$/fm irrespective of the employed parton distribution. This corresponds to $\hat{q}_{0}=0.10 \pm 0.01$ GeV$^{2}$/fm which is in agreement with the value of transport coefficient $\hat{q}_{0}=0.07 - 0.09$ GeV$^{2}$/fm used in Ref.~\cite{Arleo:2018zjw} and solely determined from the $J/\psi$ data in fully coherent energy loss regime~\cite{Arleo:2012rs}. It may also be noted that for all our calculations we have employed central set parton densities which uses most probable values of the opted EPPS21 NLO nPDF parameters. However, EPPS21, like all other nPDFs has substantial uncertainty bands reflecting how well the parameters are determined from the underlying data. Unlike its previous variants (EPS09 or EPPS16), EPPS21 nPDF has also explored the uncertainties of the nuclear PDFs due to baseline free proton PDFs via error propagation within the Hessian framework. The EPPS21 nPDF package~\cite{Eskola:2021nhw} thus features total 106 error sets of which 1 - 48 represent uncertainties due to nuclear modifications and 49 - 106 are due to uncertainties in the baseline proton PDFs CT18ANLO. As the nuclear modification in parton distributions also lead to suppression of DY production cross section in p+A collisions, it will be interesting to examine the effect of PDF uncertainties on the initial state energy loss extracted from the data. For this purpose we consider five representative nPDF error sets namely 1, 27, 47, 51 $\&$ 106 and estimate the mean quark energy loss within BDMPS formalism and the resulting quark transport coefficient $\hat{q}$ in cold nuclear matter using both E866 and E906 DY data sets. At 800 GeV the mean value of $\hat{q}$ changes around $ 20 - 30 \%$ for error sets 1, 27 $\&$ 47 accounting for uncertainties in nuclear modification and less than $4 \%$ for error sets 51 $\&$ 106 corresponding to uncertainties in the baseline free proton PDF. Uncertainties associated with quark densities inside a free proton is much lesser than that inside a bound nucleon.  However, the effect of uncertainties in quark density distribution in general is minimal on the extracted mean energy loss and resulting $\hat{q}$ from E906 data at 120 GeV which is less than $8 \%$ for error sets accounting unceratinties in nuclear modification and less than $3 \%$ while considering uncertainty coming from baseline free proton PDF. This re-establishes the fact that the DY measurement at lower collision energies probe a kinematic domain where nuclear modification of parton densities is expected to be much less important than initial state energy loss. Similar observation holds true for BH scheme of quark energy loss as well.


Similarly, the mean value of specific energy loss as obtained from the BH model can be incorporated into the NVZ model to estimate the average $X_{0}$, as both the formalisms assume a linear path length dependence of the initial state quark energy loss in cold nuclear matter. Within NVZ model, the rate of energy loss is parametrized as $-dE/dx = E/X_{0}$. Considering $<x_{1}> \simeq 0.64$ for the DY events collected by E906 experiment, the average energy of the incident quark comes out to be 76.8 GeV. Taking $<-dE/dx> \approx 0.47$ \linebreak GeV/fm from Table~\ref{tab:fitresult} the mean quark radiation length comes out to be around $X_{0} \simeq 163$ fm. The estimated value is in line with the calculations presented in~\cite{Neufeld:2010dz} and represents the shortest radiation length in nature.

To further test the path-length dependence of energy loss effect from the existing measurements we check the behavior of $R_{pA}^{DY} (A/C)$ at high $x_{F}$ as a function of linear size of the target nucleus measured in terms of $A^{1/3}$. $R_{pA}^{DY} \equiv {{\sigma_{pA}/A} \over {\sigma_{pC}/12}}$ is the ratio of the per nucleon DY production cross sections in p+A and p+C collisions. Considering the $R_{pA}^{DY}$ at $x_{F}=0.73$, the highest available $x_{F}$ bin of W/C and Fe/C data, where the suppression due to energy loss is most significant, we plot in Fig.~\ref{fig:E906_LA} the attenuation of $R_{pA}^{DY}$ as a function of $A^{1/3}$. As evident, the result fits the description of both linear and quadratic path length dependencies with equal acumen. With only three data points currently available with their associated uncertainties no conclusion can be made yet. An additional data point can be obtained from the future analysis evaluating the cross section ratio between solid targets and deuterium, which might help to further constraint the $A^{1/3}$ dependency of the energy loss.


\par
As mentioned earlier, the NA50 collaboration at CERN SPS has measured the DY production cross sections in 400 GeV and 450 GeV p+A collisions for a variety of nuclear targets. However, instead of differential analysis, the inclusive cross sections over the entire phase space window were reported. Data do not show any significant suppression with increasing size (A) of the target nucleus. For completeness we compare the integrated DY cross section ratios with our model calculations at both the energies. The results are displayed in Fig.~\ref{fig:NA50}. The theoretical curves correspond to three different scenarios namely the nuclear shadowing of target quarks without any beam quark energy loss and target shadowing and incoming quark energy loss modeled using BH or BDMPS formalism. In case of energy loss, we have taken the corresponding parameter ($\alpha$ or $\beta$) value same as the mean value at 120 GeV from Table~\ref{tab:fitresult}. Any energy loss parametrization as well as only nuclear modification of parton densities without any energy loss effect all can describe the inclusive DY data within error bars. Inclusive DY cross sections even measured for a variety of targets are thus not suitable to investigate the initial state energy loss effects. \\

Before proceeding further it would be important to discuss the limitations of the present work. Throughout our calculations we have used LO DY cross sections. Higher order contributions are accounted through the phenomenological $K$-factor. At NLO level both Compton scattering and annihilation graphs have contributions to the DY process. As we have analyzed the ratio of DY production cross sections, we assume the effect of NLO contributions would cancel. Also we have neglected the fluctuations in the energy loss of incoming beam quarks and instead use a mean initial state energy loss for both BH and BDMPS formalisms. A more rigorous analysis requires to account for a realistic distribution of $P_{q}(\epsilon)$, the probability for an incoming quark to lose $\epsilon$ fraction of its energy due to multiple gluon emission. Using mean energy loss instead of full convolution over $P_{q}(\epsilon)$ in computing DY cross section over predicts the suppression at large $x_{F}$. Typical quark energy loss that contributes to the observed suppression are smaller than the mean energy loss $\Delta{E}$ and hence our extracted parameters $\alpha$ or $\beta$ can be considered as the upper limits. We plan to investigate the effect of fluctuation in a future work. 

We close this section with the discussion on the prospect of determining the stopping power of cold nuclear matter in the upcoming fixed target proton induced DY reactions at various accelerator facilities. 50 GeV proton beam will be available from Japan Proton Accelerator Research (J-PARC) complex, whereas di-muon production in p+A collisions will be studied by the NA60+ spectrometer at SPS within the energy range $40 - 160$ GeV. The CBM experiment at FAIR SIS100 will also measure di-muons in 30 GeV p+A collisions. Such low energy collisions offer unique opportunity to disentangle initial-state energy loss effect from the nuclear shadowing. Studying DY process in these experiments in a region of phase space where shadowing effects are small will help in precise determination of quark energy loss and its kinematic dependence. The ratio of DY production cross sections, $R_{pA}^{DY} (A/Be)$ evaluated at $x_{F}=0.8$ in the dimuon mass range $1.5<m_{\mu\mu}<2.9$ GeV$/\it{c^{2}}$ as a function of the linear size of the target nucleus ($A^{1/3}$), as obtained from our model calculations are displayed in Fig.~\ref{fig:Prospects} for 30 GeV, 50 GeV and 80 GeV p+A collisions. Our model predictions indicate a small attenuation of $R_{pA}^{DY}$ at high $x_{F}$ that eventually becomes weaker with decreasing beam energy in presence of nuclear shadowing effects alone. On contrary, a significant suppression of DY di-lepton production cross section in p+A reactions is anticipated due to incoming quark energy loss with increasing linear size ($\propto A^{1/3}$) of the target nucleus. Lower is beam energy steeper attenuation of $R_{pA}^{DY}$ at forward $x_{F}$ induced by energy loss is observed. Two models of energy loss generate distinct suppression patterns and the difference increases with decreasing beam energy. Collection of high statistics DY data at large $x_{F}$ (rapidity) by these experiments will thus be useful to better constraint the quark energy loss in cold nuclear matter along with a definitive confirmation of its kinematic dependence. However, the major challenge in performing these experimental measurements is the detection of a statistically significant suppression pattern, owing to extremely low production high mass DY events at these low energies. 

\section{Summary}
Unambiguous determination of stopping power of cold nuclear matter for energetic partons is one of the most important issues in the theory and phenomenology of heavy ion reactions at relativistic energies. The DY reaction in proton induced collisions has been identified as a suitable tool to estimate the initial state quark energy loss effects. In these reactions the fast parton from the projectile hadron propagates through the cold nuclear matter of the target nucleus and can suffer energy loss due to multiple soft collisions before the annihilation takes place. As the produced di-muons do not interact with the surrounding nuclear medium via strong interaction, the final state effects for such reactions can be ignored. This makes DY collisions an ideal tool to probe the initial state energy loss effects in nuclear collisions. Due to the energy loss effect the energy of the incident beam parton just before the high $Q^{2}$ annihilation would be different from its initial value. This shift in energy can be equivalently expressed as a change in Bjorken-$x$, the fraction of the hadron momentum carried by the parton participating in the annihilation process. An observable signature of initial state energy loss would thus be the modification of the di-muon spectrum with respect to Bjorken-$x$ or $x_{F}$. As the incoming quarks presumably suffer larger energy losses in heavier nuclei, it can be estimated by measuring the nuclear dependence of the DY differential cross sections in p+A collisions. However, presence of strong shadowing effects makes the extraction of magnitude of energy loss from the DY data highly uncertain, though they have completely different physics origin. 
 The DY data measured by E866 experiment in 800 GeV p+A collisions has been found to be described both in presence and absence of initial state energy loss effects once the nuclear modification of parton densities are taken into account. The magnitude of energy loss extracted from the data heavily depends on the underlying nPDF scheme employed to incorporate the shadowing effects inside the target. The situation improves as the DY data in 120 GeV p+A collisions from Fermilab E906 experiment is now becoming available. Its low centre of mass energy per nucleon pair $\sqrt{s_{NN}} = 15$ GeV and selected di-muon mass range allows to probe a kinematic domain where the shadowing effects are minimal. The observed suppression in the DY dilepton production cross section with increasing $x_{F}$ can then be unambiguously associated with the presence of energy loss effects. However, the present data do not offer the possibility to determine the path length dependence of the mean energy loss. The BH and BDMPS models respectively having linear and quadratic dependence of the quark energy loss on nuclear path length can describe the data within errors. The situation might improve in near future with the availability of DY cross section ratios between solid targets and deuterium analyzed from full recorded statistics with improved statistical and systematic uncertainties. The other opportunity to measure the quark energy loss in cold nuclear matter is to study DY production in lower energy fixed target p+A collisions as will be available from different upcoming accelerator facilities. With lower proton beam energy compared to E906, our model calculations predict a steeper attenuation of $R_{pA}$ at larger $x_{F}$ due to initial energy loss effects prior to a large $Q^{2}$ scattering. 
As the beam energy decreases, the difference in the suppression pattern becomes more pronounced, presenting an opportunity for experimental distinction using high-precision DY data from various nuclear targets.

\begin{acknowledgements}
We are thankful to F. Arleo for useful discussions. SKD and PPB acknowledge the support from DAE-BRNS, India, Project No. 57/14/02/2021-BRNS.
\end{acknowledgements}


\begin{thebibliography}{50}

\bibitem{Bjorken:1982tu}
J.~D.~Bjorken,
FERMILAB-PUB-82-059-THY

\bibitem{Lee:2013bka}
Y.~J.~Lee,
J. Phys. Conf. Ser. \textbf{446}, 012001 (2013)
doi:10.1088/1742-6596/446/1/012001

\bibitem{Cao:2020wlm}
S.~Cao and X.~N.~Wang,
Rept. Prog. Phys. \textbf{84}, no.2, 024301 (2021)
doi:10.1088/1361-6633/abc22b
[arXiv:2002.04028 [hep-ph]].

\bibitem{Cunqueiro:2021wls}
L.~Cunqueiro and A.~M.~Sickles,
Prog. Part. Nucl. Phys. \textbf{124}, 103940 (2022)
doi:10.1016/j.ppnp.2022.103940
[arXiv:2110.14490 [nucl-ex]].

\bibitem{Drell:1970wh}
 S. D. Drell and T. M. Yan, Phys. Rev. Lett. \textbf{25}, 316-320 (1970)

\bibitem{Arleo:2002ki}
F.~Arleo,
Nucl. Phys. A \textbf{715}, 899-902 (2003)
doi:10.1016/S0375-9474(02)01537-3
[arXiv:hep-ph/0210105 [hep-ph]]

\bibitem{Arleo:2003jz}
F.~Arleo,
Eur. Phys. J. C \textbf{30}, 213-221 (2003)
doi:10.1140/epjc/s2003-01289-x
[arXiv:hep-ph/0306235 [hep-ph]].
 
 \bibitem{Song:2010zza}
L.~H.~Song and C.~G.~Duan,
Phys. Rev. C \textbf{81}, 035207 (2010)
doi:10.1103/PhysRevC.81.035207
[arXiv:1109.3836 [hep-ph
]].
\bibitem{Vitev:2007ve}
I.~Vitev,
Phys. Rev. C \textbf{75}, 064906 (2007)
doi:10.1103/PhysRevC.75.064906
[arXiv:hep-ph/0703002 [hep-ph]].

\bibitem{Alde:1990im}
D.~M.~Alde, H.~W.~Baer, T.~A.~Carey, G.~T.~Garvey, A.~Klein, C.~Lee, M.~J.~Leitch, J.~W.~Lillberg, P.~L.~McGaughey and C.~S.~Mishra, \textit{et al.}
Phys. Rev. Lett. \textbf{64}, 2479-2482 (1990)
doi:10.1103/PhysRevLett.64.2479

 \bibitem{NuSea:1999egr}
M.~A.~Vasilev \textit{et al.} [NuSea],
Phys. Rev. Lett. \textbf{83}, 2304-2307 (1999)
doi:10.1103/PhysRevLett.83.2304
[arXiv:hep-ex/9906010 [hep-ex]].

\bibitem{Eskola:1998iy}
K.~J.~Eskola, V.~J.~Kolhinen and P.~V.~Ruuskanen,
Nucl. Phys. B \textbf{535}, 351-371 (1998)
doi:10.1016/S0550-3213(98)00589-6
[arXiv:hep-ph/9802350 [hep-ph]].



	\bibitem{Gavin:1991qk}
S.~Gavin and J.~Milana,
Phys. Rev. Lett. \textbf{68}, 1834-1837 (1992)



  
\bibitem{Brodsky:1992nq}
S.~J.~Brodsky and P.~Hoyer,
Phys. Lett. B \textbf{298}, 165-170 (1993)
doi:10.1016/0370-2693(93)91724-2
[arXiv:hep-ph/9210262 [hep-ph]].
 
 \bibitem{Baier:1996sk}
  R. Baier {\textit {et. al.}},
Nucl. Phys. B \textbf{484}, 265 (1997)

\bibitem{Johnson:2001xfa}
M.~B.~Johnson, B.~Z.~Kopeliovich, I.~K.~Potashnikova, P.~L.~McGaughey, J.~M.~Moss, J.~C.~Peng, G.~Garvey, M.~Leitch, C.~N.~Brown and D.~M.~Kaplan,
Phys. Rev. C \textbf{65}, 025203 (2002)
doi:10.1103/PhysRevC.65.025203
[arXiv:hep-ph/0105195 [hep-ph]].

\bibitem{FNALE772:2000fmo}
M.~B.~Johnson \textit{et al.} [FNAL E772],
Phys. Rev. Lett. \textbf{86}, 4483-4487 (2001)
doi:10.1103/PhysRevLett.86.4483
[arXiv:hep-ex/0010051 [hep-ex]].

\bibitem{Garvey:2002sn}
G.~T.~Garvey and J.~C.~Peng,
Phys. Rev. Lett. \textbf{90}, 092302 (2003)
doi:10.1103/PhysRevLett.90.092302
[arXiv:hep-ph/0208145 [hep-ph]].








\bibitem{Arleo:2002ph}
F.~Arleo,
Phys. Lett. B \textbf{532}, 231-239 (2002)
doi:10.1016/S0370-2693(02)01539-3
[arXiv:hep-ph/0201066 [hep-ph]].

\bibitem{NA3:1983ltt}
J.~Badier \textit{et al.} [NA3],
Z. Phys. C \textbf{20}, 101 (1983)
doi:10.1007/BF01573213


\bibitem{Duan:2005wj}
C.~G.~Duan, G.~L.~Li, L.~H.~Song and S.~H.~Wang,
Eur. Phys. J. C \textbf{39}, 179-184 (2005)
doi:10.1140/epjc/s2004-02074-1
[arXiv:hep-ph/0601188 [hep-ph]].


\bibitem{Duan:2008qt}
C.~G.~Duan, N.~Liu and G.~L.~Li,
Phys. Rev. C \textbf{79}, 048201 (2009)
doi:10.1103/PhysRevC.79.048201
[arXiv:0811.0675 [hep-ph]]


\bibitem{Song:2012zz}
L.~H.~Song, C.~G.~Duan and N.~Liu,
Phys. Lett. B \textbf{708}, 68-74 (2012)
doi:10.1016/j.physletb.2012.01.019
[arXiv:1206.3815 [hep-ph]].

\bibitem{Song:2017wuh}
L.~H.~Song and L.~W.~Yan,
Phys. Rev. C \textbf{96}, no.4, 045203 (2017)
doi:10.1103/PhysRevC.96.045203

\bibitem{Neufeld:2010dz}
R.~B.~Neufeld, I.~Vitev and B.~W.~Zhang,
Phys. Lett. B \textbf{704}, 590-595 (2011)
doi:10.1016/j.physletb.2011.09.045
[arXiv:1010.3708 [hep-ph]].

\bibitem{Ayuso:2020zht}
C.~Ayuso, doi:10.2172/1637630


\bibitem{Arleo:2018zjw}
F.~Arleo, C.~J.~Na\"\i{}m and S.~Platchkov,
JHEP \textbf{01}, 129 (2019)
doi:10.1007/JHEP01(2019)129
[arXiv:1810.05120 [hep-ph]].

\bibitem{Song:2020vhy}
L.~H.~Song, S.~F.~Xin and Y.~J.~Zhang,
J. Phys. G \textbf{47}, no.5, 055002 (2020)
doi:10.1088/1361-6471/ab72b1
[arXiv:2003.10610 [hep-ph]].

\bibitem{Song:2021mzt}
L.~H.~Song, P.~Q.~Wang and Y.~J.~Zhang,
Chin. Phys. C \textbf{45}, no.4, 041005 (2021)
doi:10.1088/1674-1137/abe110
[arXiv:2101.11248 [hep-ph]].

\bibitem{NA10:1987hho}
P.~Bordalo \textit{et al.} [NA10],
Phys. Lett. B \textbf{193}, 368 (1987)
doi:10.1016/0370-2693(87)91253-6

\bibitem{Xu:2022usl}
W.~J.~Xu, T.~X.~Bai and C.~G.~Duan,
Chin. Phys. C \textbf{47}, no.4, 043110 (2023)
doi:10.1088/1674-1137/acb8a4
[arXiv:2201.08577 [hep-ph]].

\bibitem{Lin:2017}
P. J. Lin, 
https://lss.fnal.gov/archive/thesis/2000/fermilab-thesis-2017-18.pdf,
Ph.D. thesis, Colorado U, (2017)


\bibitem{Nagamiya:2006en}
S.~Nagamiya,
Nucl. Phys. A \textbf{774}, 895-898 (2006)
doi:10.1016/j.nuclphysa.2006.06.160


\bibitem{Falco:2023hwb}
A.~D.~Falco [NA60+],
EPJ Web Conf. \textbf{276}, 05005 (2023)
doi:10.1051/epjconf/202327605005

\bibitem{Alocco:2024hvm}
G.~Alocco [NA60+],
EPJ Web Conf. \textbf{296}, 08005 (2024)
doi:10.1051/epjconf/202429608005

\bibitem{Bhaduri:2022cql}
P.~P.~Bhaduri [CBM],
PoS \textbf{CPOD2021}, 031 (2022)
doi:10.22323/1.400.0031


\bibitem{Gavin:1995ch}
S.~Gavin, R.~Kauffman, S.~Gupta, P.~V.~Ruuskanen, D.~K.~Srivastava and R.~L.~Thews,
Int. J. Mod. Phys. A \textbf{10}, 2961-2998 (1995)
doi:10.1142/S0217751X9500142X
[arXiv:hep-ph/9502372 [hep-ph]]

\bibitem{Collins:1989gx}
J.~C.~Collins, D.~E.~Soper and G.~F.~Sterman,
Adv. Ser. Direct. High Energy Phys. \textbf{5}, 1-91 (1989)
doi:10.1142/9789814503266\_0001
[arXiv:hep-ph/0409313 [hep-ph]].

\bibitem{deFlorian:2011fp}
D.~de Florian, R.~Sassot, P.~Zurita and M.~Stratmann,
Phys. Rev. D \textbf{85}, 074028 (2012)
doi:10.1103/PhysRevD.85.074028
[arXiv:1112.6324 [hep-ph]].



\bibitem{Kovarik:2015cma}
K.~Kovarik, A.~Kusina, T.~Jezo, D.~B.~Clark, C.~Keppel, F.~Lyonnet, J.~G.~Morfin, F.~I.~Olness, J.~F.~Owens and I.~Schienbein, \textit{et al.}
Phys. Rev. D \textbf{93}, no.8, 085037 (2016)
doi:10.1103/PhysRevD.93.085037
[arXiv:1509.00792 [hep-ph]].



\bibitem{Eskola:2016oht}
K.~J.~Eskola, P.~Paakkinen, H.~Paukkunen and C.~A.~Salgado,
Eur. Phys. J. C \textbf{77}, no.3, 163 (2017)
doi:10.1140/epjc/s10052-017-4725-9
[arXiv:1612.05741 [hep-ph]].





\bibitem{Eskola:2021nhw}
K.~J.~Eskola, P.~Paakkinen, H.~Paukkunen and C.~A.~Salgado,
Eur. Phys. J. C \textbf{82}, no.5, 413 (2022)
doi:10.1140/epjc/s10052-022-10359-0
[arXiv:2112.12462 [hep-ph]].

\bibitem{NA50:2006rdp}
B.~Alessandro \textit{et al.} [NA50],
Eur. Phys. J. C \textbf{48}, 329 (2006)
doi:10.1140/epjc/s10052-006-0079-4
[arXiv:nucl-ex/0612012 [nucl-ex]].


\bibitem{NA50:2003pvd}
B.~Alessandro \textit{et al.} [NA50],
Phys. Lett. B \textbf{553}, 167-178 (2003)
doi:10.1016/S0370-2693(02)03265-3





\bibitem{Eskola:2009uj}
K.~J.~Eskola, H.~Paukkunen and C.~A.~Salgado,
JHEP \textbf{04}, 065 (2009)
doi:10.1088/1126-6708/2009/04/065
[arXiv:0902.4154 [hep-ph]].

\bibitem{Xie:2021equ}
K.~Xie \textit{et al.} [CTEQ-TEA],
Phys. Rev. D \textbf{105}, no.5, 054006 (2022)
doi:10.1103/PhysRevD.105.054006
[arXiv:2106.10299 [hep-ph]].
 



\bibitem{root} V.5.34/32, CERN ROOT, 2015, http://root.cern.ch

\bibitem{minuit2} F. James and M. Roos, Comput. Phys. Commun. {\bf 10}, 343 (1975).

\bibitem{Arleo:2012rs}
F. Arleo and S. Peigne,
JHEP \textbf{03}, 122 (2013)
doi:10.1007/JHEP03(2013)122
[arXiv:1212.0434 [hep-ph]].



\end{thebibliography}
\end{document}